\documentclass[useAMS,usenatbib]{mn2e}
\usepackage{natbib}
\usepackage{graphicx}

\title[The effect of gas expulsion]{Possible smoking-gun evidence for initial mass segregation in re-virialized post-gas expulsion globular clusters}

\author[Haghi et al.]
{Hosein Haghi $^{1}$\thanks{
E-mail:  \mbox{haghi@iasbs.ac.ir} (HH)   
\mbox{a.hasani@iasbs.ac.ir} (AHZ);
\mbox{pavel@astro.uni-bonn.de} (PK);
\mbox{sambaran@astro.uni-bonn.de}(SB);
\mbox{h.baumgardt@uq.edu.au}(HB);
 }
, Akram Hasani Zonoozi$^{1}$, Pavel Kroupa$^2$, Sambaran Banerjee$^{2,3}$,
\newauthor
Holger Baumgardt$^4$\\\\
$^{1}$ Department of Physics, Institute for Advanced Studies in Basic Sciences (IASBS), P.O. Box 11365-9161, Zanjan, Iran\\
$^{2}$ Helmholtz-Institut f\"ur Strahlen-und Kernphysik (HISKP), Universit\"at Bonn, Rheinische Friedrich-Wilhelms-Universit\"at\\
Nussallee 14-16 D-53115 Bonn Germany\\
$^{3}$Argelander Institute f\"ur Astronomie (AIfA), Auf dem H\"ugel 71, 53121 Bonn, Germany\\
$^{4}$University of Queensland, School of Mathematics and Physics, Brisbane, QLD 4072, Australia\\
}

\begin{document}

\date{Accepted \ldots. Received \ldots; in original form \ldots}

\pagerange{\pageref{firstpage}--\pageref{lastpage}} \pubyear{2010}

\maketitle

\label{firstpage}

\maketitle

\begin{abstract}
We perform a series of direct $N$-body calculations to investigate the effect of residual gas expulsion from the gas-embedded progenitors of present-day globular clusters (GCs) on the stellar mass function (MF). Our models start either tidally filling or underfilling, and either with or without primordial mass segregation. We cover 100 Myr of the evolution of modeled clusters and show that the expulsion of residual gas from initially mass-segregated clusters leads to a significantly shallower slope of the stellar MF in the low- ($m\leq 0.50 M_\odot$) and intermediate-mass ($\simeq 0.50-0.85 M_\odot$) regime.
Therefore, the imprint of residual gas expulsion and primordial mass segregation might be visible in the present-day MF. We find that the strength of the external tidal field, as an essential parameter, influences the degree of flattening, such that a primordially mass-segregated tidally-filling cluster with $r_h/r_t\geq 0.1$ shows a strongly depleted MF in the intermediate stellar mass range. Therefore, the shape of the present-day stellar MF in this mass range probes the birth place of clusters in the Galactic environment.  We furthermore find that this flattening agrees with the observed correlation between the concentration of a cluster and its MF slope, as found by de Marchi et al.. We show that if the expansion through the residual gas expulsion in primordial mass segregated clusters is the reason for this correlation then GCs most probably formed in strongly fluctuating local tidal fields in the early proto-Milky Way potential, supporting the recent conclusion by Marks \& Kroupa.

\end{abstract}


\section{Introduction}

The study of dense stellar systems, which mainly refers to young massive star clusters, globular clusters and ultra-compact dwarf galaxies, has played an important role in developing our knowledge about the universe. Since most stars in the Galactic disc may originate in embedded star clusters, these systems can be viewed as the fundamental building blocks of galaxies to understand the origins of the properties of the galactic stellar population, such as the galaxy-wide stellar mass function \citep{Weidner05, Kroupa05}.

The internal properties  of star clusters can undergo significant changes at birth and also during the course of cluster evolution. Hence, it is important to specify to what amount the current properties of star
clusters are imprinted at the beginning of the evolution and early formation processes and to what amount  they are the result of long-term evolutionary processes (Vesperini 2010).  In fact, star formation provides the initial conditions for the long-term evolution of stellar clusters, while the follow up dynamics of stars in clusters can inform on the conditions that have dominated in the environment in which the clusters have formed.

The stars of star clusters are born more or less simultaneously within the dense cores of the same progenitor giant molecular cloud \footnote{However, many observations suggest that massive clusters appear to be composed of (at least) two distinct stellar populations (e.g., Piotto 2010)}.
Stars within an embedded cluster form following an initial mass function (IMF).  The shape of the stellar IMF is one of the main uncertainties in our understanding of star formation, i.e., the conversion of gas into stars over cosmic time. It is usually assumed that the IMF in external galaxies is the same as in the disk of the Milky Way, where most studies of resolved stellar populations showed the IMF to be invariant (Bastian et al. 2010, Kroupa et al. 2013). The IMF is also observationally found to be invariant for nearly all dense systems in which its shape could be inferred and which formed with a star formation density below a threshold near $0.1 M_{\odot}/(yr pc^3)$ and is hence referred to as the "standard" or "canonical" IMF. This poses a problem for star formation theories which predict a dependence on the environment where star formation takes place (Kroupa et al 2013).

In the course of time, the MF of stars in clusters evolves at early stages and also during the cluster secular evolution as the stellar members are removed from the system through stellar and dynamical evolution (Vesperini \& Heggie 1997, Baumgardt \& Makino 2003). Simulations show that mass segregation and the preferential loss of low-mass stars which leads to a shallower mass function should only happen after a cluster has gone into two-body relaxation-driven core-collapse which is connected to the shrinkage of the core size $r_c$ (Baumgardt \& Makino 2003). So one would expect a correlation between MF-slope, $\alpha$ ($dN/dm\propto m^{-\alpha}$, $\alpha=2.3$  being the Salpeter index), and concentration parameter ($c=\log(r_t/r_c)$), in the sense that clusters with large values of $c$ are depleted in low-mass stars (fig. 4 in Leigh et al. 2013). Here, $dN$ is the number of stars in the mass interval $m$ to $m+dm$ and $r_t$, $r_c$ are the tidal and core radius, respectively.

But, according to observational data by De Marchi, Paresce, \& Pulone (2007), and \cite{Paust10} for Galactic globular clusters the value of $\alpha$, in the mass range 0.3-0.8 $M_{\odot}$, correlates surprisingly with their concentration parameter, such that the less concentrated clusters are typically but not exclusively more strongly depleted in low mass stars. This trend can not be understood to be the result of long-term evolution alone and  at least some initial correlation between the initial concentration and the total cluster mass is required (Leigh et al 2013).

The 'De Marchi relation' could be interpreted as a first evidence for a non-universality of the IMF at low masses, but Baumgardt et al (2008a) showed that such a correlation can be understood if clusters are born initially mass segregated with a canonical IMF, but are filling their tidal radii.  The difficulty of this scenario is that the cluster would have to form with knowledge of its mass-dependent tidal radius at different locations in a galaxy. It is unclear how star formation can arrange this. Marks et al. (2008), however, demonstrated, using N-body integrations of young, gas embedded compact clusters, that the loss of the primordial residual-gas from initially mass-segregated clusters in a range of tidal field strenghts starting with a canonical IMF  and containing unresolved binaries reasonably reproduces the relation between the slope of the actual mass function and the concentration of globular clusters within the observational limits.

The flattening of the MF-slope driven by a violent early phase of gas-expulsion of an embedded cluster with primordial mass segregation can also be a possible explanation for the observed flattened mass function of some remote halo globular clusters like for example, Palomar 4 and Palomar 14 (Zonoozi et al. 2011, 2014)

In this paper, our aim is to consider the effect of primordial residual-gas expulsion on the early evolution of globular clusters (covering the first 100 Myr of their evolution) and it's influence on the stellar MF-slope  and degree of mass segregation for a wide range of stellar masses, varying the star formation efficiency, and the strength of the external tidal field. Contrary to the otherwise identical models studied by Marks et al. (2008), we here use a fully realistic distribution of stellar masses. Our paper is organized as follows: In Sec. 2 we discuss the star cluster reaction to residual-gas expulsion. In Sec. 3 we describe the initial set-up of the N-body models and in Sec. 4 we present the main results of the simulations. Finally Sec. 5 consists of the conclusions and a discussion.

\section{The reaction of star clusters to residual-gas expulsion}

The fraction of gas that is converted into stars within the volume of the embedded cluster is defined as the star formation efficiency (SFE),
\begin{equation}
 \varepsilon = \frac{M_{ecl}}{M_{ecl}+M_{gas}},
\end{equation}
where $M_{ecl}$ is the stellar mass of the embedded cluster and $M_{gas}$ is the residual gas within the cluster. A large SFE leads to more stars and therefore also to more massive stars with stronger winds and radiation and there is less gas to remove. The observationally determined SFE is usually smaller than $\varepsilon \simeq 40\%$ (Lada \& Lada 2003) so most of the mass of the progenitor cloud remains in the form of residual gas to be expelled.

In young, still gas-embedded star clusters, the left-over gas from star formation is lost over a time which is determined by feedback from massive stars through UV radiation and massive stellar winds from OB stars or supernova explosions. Thus, star clusters will become super-virial and their further dynamical evolution will be strongly affected by the gas loss.

Recent observations indicate that proto-stars form in molecular clouds as complex and filamentary substructures. However, as shown in Banerjee \& Kroupa (2015), if a young star cluster has to form out of them, they must fall in the potential well of the molecular cloud and form a near-spherical and nearly virialized cluster within a few dynamical times, which is short (or similar) compared to the typical gas dispersal time that we take here ($\approx 0.6$ Myr). Observational evidence for such a monolithic formation of individual massive cluster rather than a hierarchical formation of successively merging of small-clusters is provided by the $> 5\times 10^6 M_{\odot}$, and $< 24$ pc sized molecular cloud formed in the Antennae merging galaxies by Johnson et al. (2015). Hence, our initial conditions, that comprise virialized embedded clusters, are plausible.  As suggested by some recent theoretical studies involving hydrodynamic calculations, e.g., Dale et al. (2015, 2012), the locations at which the protostars (or sink particles) are formed, the star formation efficiency is nearly $100\%$ and so the removal of gas is less efficient in further dynamics, arguing against  gas-expulsion as being a relevant physical process that governs the long-term evolution of a cluster. This is likely to be a result of including only radiative feedback and ignoring magnetic fields as well as performing simulations in which the sink particles are significantly more massive than individual stars. However, according to the high resolution hydrodynamic calculations by Machida \& Maatsumoto (2012) and  Bate et al. (2014), and direct observation (Lada \& Lada 2003) only small fraction of gas convert into stars.

Based on the hydrodynamical simulations without feedback, \cite{Kruijssen12} also claimed that the SFE is higher in the regions where stars have formed, and hence the long-term evolution of star clusters is only weakly affected by gas removal. They argued that this is because of the low gas fractions within the pre-cluster cloud core, caused by the accretion of gas on to the stars. In this case it however needs to be explained how the outer parts of the gas cloud are expelled without heating and pushing out the inner gas contents of the pre-cluster cloud core. It also is difficult to explain ultracompact HII regions if gas can simply be accreted onto the stars despite their luminosity. In the short-lived UCHII phase, the gas is heated significantly (up to $10^4$ K or more) in the immediate surrounding of the stars (within $0.05$ pc), which then leads to an outflow of the residual gas on larger scales. 

However, in the simulations by \cite{Dale12} including the influence of photoionizing radiation from O-type stars it is shown that the global SFE is between 20-30\% and agrees with what is observed in star-forming regions and embedded clusters which suggests that the star formation efficiency can at most be $\varepsilon \approx 30\%$ (see, e.g., Lada \& Lada 2003), although on the scale of whole molecular clouds the star formation efficiency is at most a few percent (Murray 2011). High-resolution mapping in star-forming regions has uncovered that outflows, even from low-mass proto-stars, are very powerful in destroying and clearing cloud material away from the star-forming region on sub-ps scales (e.g., Bally et al. 2011, 2012, McGroarty et al. 2004, Frank et al. 2014). 

In addition, the dynamical state of the stars at the onset of gas expulsion might be a key factor in determining the effects of gas expulsion \citep{Verschueren89, Goodwin09}. For example, by means of N-body simulations of highly non-equilibrium star clusters in fixed background potential (as a simple model of gas potential), Smith et al. (2011) found that  the long-term dynamical evolution of star clusters is affected by the initial stellar distribution and velocity dispersion far more than by gas expulsion. They have shown that initially cool clusters are more able to survive gas expulsion, weakening the dependency of survival to gas expulsion on the initial SFE compared to a virialised initial dynamical state of the clusters. Such an ansatz however requires all stars to have formed in some volume synchronically on a time-scale much shorter than the dynamical time-scale of this volume, which appears to be an artificial initial state. Further discussion of this issue is available in Kroupa (2005). Observations of UCHII regions, of outflows and magnetohydrodynamical simulations of star formation show that the SFE does not surpass about 20-30\% on a sub-pc scale because matter is blown out in outflows launched by the magnetic fields.  This questions the above mentioned results according to which gas-removal is not important.

The main argument here is that actual proto-star scale high-resolution MHD calculations (e.g., Machida \& Maatsumoto 2012, Bate et al. 2014) show that the  interplay between gas accretion onto the protostar and magnetically driven outflow results in star formation efficiencies up to $\approx 30\%$ in the neighborhood of proto-stars, which is a well sub-pc scale. This would imply the overall SFE in a dense gas clump (sub-pc to pc scale) would be less than 30\%, including its innermost regions. Such proto-star scale calculations are the only available ones that can track the mass accretion and outflow (driven by radiation and magnetic field) from individual proto-stars. Other more massive SPH simulations, that intend to form star clusters, either exclude all feedback or include only partial feedback. The SFE from these calculations can never be reliable. The Dale et al. (2015) calculations that go up to globular cluster masses had their sink particle representing a sub-cluster itself, i.e., their sink particles themselves are of pc scale. Of course, with a 30\% SFE all over the clump (the SFE can, in practice, be smaller in the outer regions), gas expulsion should have a significant impact on the cluster. This is the initial star-to-gas ratio that we plausibly adopt in our computations.

We assume that the gas expulsion starts  after the embedded phase at a certain time, $\tau_{D}$, which is set equal to  $0.6$  Myr and corresponds to the UCHII phase. After this time, it is assumed that the mass of gas decreases exponentially on the gas-expulsion characteristic time scale $\tau_{GE}$, over which the natal gas is removed from the cluster. So the time-varying total mass in gas is given by

\begin{equation}
 M_{gas}(t) = M_{gas}(0)$ $ e^{-(t-\tau_{D})/\tau_{GE}}.
 \end{equation}

The impact of the early gas expulsion on the early dynamical evolution of star clusters has been investigated by Baumgardt \& Kroupa (2007) using a large set of N-body integration. They showed that the survival rate and final properties of star clusters are strongly influenced by the details of the expulsion of the residual-gas that did not form stars,  depending on the parameter values for the gas-expulsion time scale, $\tau_{GE}$, the star formation efficiency, $\varepsilon$, and the ratio of the initial half-mass radius to tidal radius. They also found that if the tidal field is weak ($r_h/r_t\leq0.05$, $r_h$ being the half-mass radius) and the gas is removed slowly ($\tau_{GE}\gg t_{cross}$, $t_{cross}$ being the crossing time defined here as $t_{cross}=2r_h/\sigma$, where $\sigma$ is the 3D velocity dispersion in the cluster), star clusters can survive even if the SFE is as low as $10\%$. It is shown here that the gas-free stellar system (after gas expulsion) expands violently and loses a fraction of its stars depending on its initial mass and concentration if $\varepsilon\approx 1/3$ and $\tau_{GE}$ is the thermal gas removal timescale. We take $\tau_{GE}=\frac{r_h}{10 km/s}$, where 10 km/s is the approximate sound velocity of the ionized gas.  Studying the dynamical effect of gas dispersal by adopting a time-varying background potential mimicking the influence of the gas with the above values for $\tau_{D}$ $\tau_{GE}$ and $\varepsilon$  is used by Banerjee \& Kroupa (2013, 2014, 2015) to explain well observed very young massive clusters such as R136 and the NGC 3603, and the ONC is likewise well accounted for using these values as a currently expanding population leading to a Pleiades-like open cluster in $\simeq 100 Myr$ (Kroupa et al. 2001). The conformity of these values to account for observed clusters of masses $10^3$, $10^4$, and $10^5 M_{\odot}$ suggests a common physical process which acts in their emergence from their embedded stage. 


\begin{table*}
\begin{minipage}{144mm}
\centering \caption{The initial and the final parameters of the N-body runs after t=100 Myr evolution. Column~1 gives the model name, in which the first two digit numbers after `rh' denote the initial half-mass radius in parsec, the last two digit numbers denote the percentage of the star formation efficiency (sfe), and the degree of mass segregation is written after the letter 'S'. The initial mass of all models is $M=10^5 M_\odot$. Column 2 gives the initial 3D half-mass radius. Columns 3,  4 and 5 contain the adopted star formation efficiency, gas expulsion timescale and initial degree of segregation, respectively.  Column 6 is the initial filling factor, where $r_{t0}= 54.3$ pc for all models. Note that the gas mass is not included in the calculation of the tidal radius.  The final slope of the mass function  after 100 Myr of evolution in the mass range $0.50\leq \frac{m}{M\odot} \leq0.85$ inside the tidal radius is presented in column 7, the initial canonical value being 2.3. The following columns give the final mass, the 3D half-mass radius  of the simulated star clusters after 100 Myr of evolution,  the final concentration $c = \log_{10}(r_t/r_c)$, as well as the two-body relaxation time of the star clusters after 100 Myr of evolution. }
 \begin{tabular}{cccccccccccc}
 \hline
 Model &  $r_{h0}$  &  $\varepsilon$  & $\tau_{GE}$  &  $S$ & $(r_h/r_t)_0$  & $\alpha_{2}$  & $M^{f}_{r<r_t}$& $r_{hf}$& $c$& $T_{rh}$ \\
       &     [pc]            & &    [Myr]  &&&   $[0.50 - 0.85] M_\odot$   &[$M_\odot$]& [pc]&&[Gyr] \\
 \hline (1)&(2)&(3)&(4)&(5)&(6) & (7) & (8)&(9)&(10)&(11)\\
  \hline
        rh10-S0-sfe25         &10 &0.25   &1  &0.0   &0.18&  2.16 &36741&22.3&1.38&13.14\\
        rh1-S0.9-sfe33        &1  &0.33   &0.1&0.9   &0.02&  2.27 &67392&4.4 &2.40&1.81\\
        rh1-S0.9-sfe25        &1  &0.25   &0.1&0.9   &0.02&  2.21 &65016&5.2 &2.10&1.50\\
        rh3-S0.9-sfe25        &3  &0.25   &0.1&0.9   &0.05&  1.85 &34930&12.9 &1.55&4.40\\
        rh5-S0.9-sfe25        &5  &0.25   &1  &0.9   &0.09&  1.96 &54274&18.6 &1.61&9.36\\
        rh7-S0.9-sfe25        &7  &0.25   &1  &0.9   &0.13&  1.45 &41716&21.6 &1.34&11.02\\
        rh10-S0.9-sfe33       &10 &0.33   &1  &0.9   &0.18&  1.08 &31486&22.6 &1.195&11.81\\
        rh10-S0.9-sfe25       &10 &0.25   &1  &0.9   &0.18&  0.81 &20060&21.5 &1.178&10.08\\
        rh10-S0.9-NoGE        &10 &1.0   &--  &0.9    &0.18&  2.28 &66038&16.2 &1.70&8.84\\
        rh1-S0-NoGE           &1  &1.0   &--  &0.0    &0.02&  2.29 &73827&1.9  &2.65&10.10\\
        rh3-S0-NoGE           &3  &1.0  &--  &0.0    &0.05&  2.27 &74379&4.3  &2.44&12.40\\
        rh5-S0-NoGE           &5  &1.0   &--  &0.0    &0.09&  2.30 &72668&7.0  &2.42&8.01\\
        rh10-S0-NoGE          &10 &1.0 &--  &0.0    &0.18&  2.22 &71039&12.7 &1.84&7.50\\
        rh20-S0-NoGE          &20 &1.0  &--  &0.0    &0.36&  2.26 &55029&22.8 &1.67&6.30\\
        rh30-S0-NoGE          &30 &1.0 &--  &0.0    &0.54&  2.34 &10864&35.0 &0.86&3.40\\

        \hline
 \end{tabular}
 \label{tab_regular}
 \end{minipage}
 \end{table*}

\begin{figure*}
\begin{center}
\includegraphics[width=150mm]{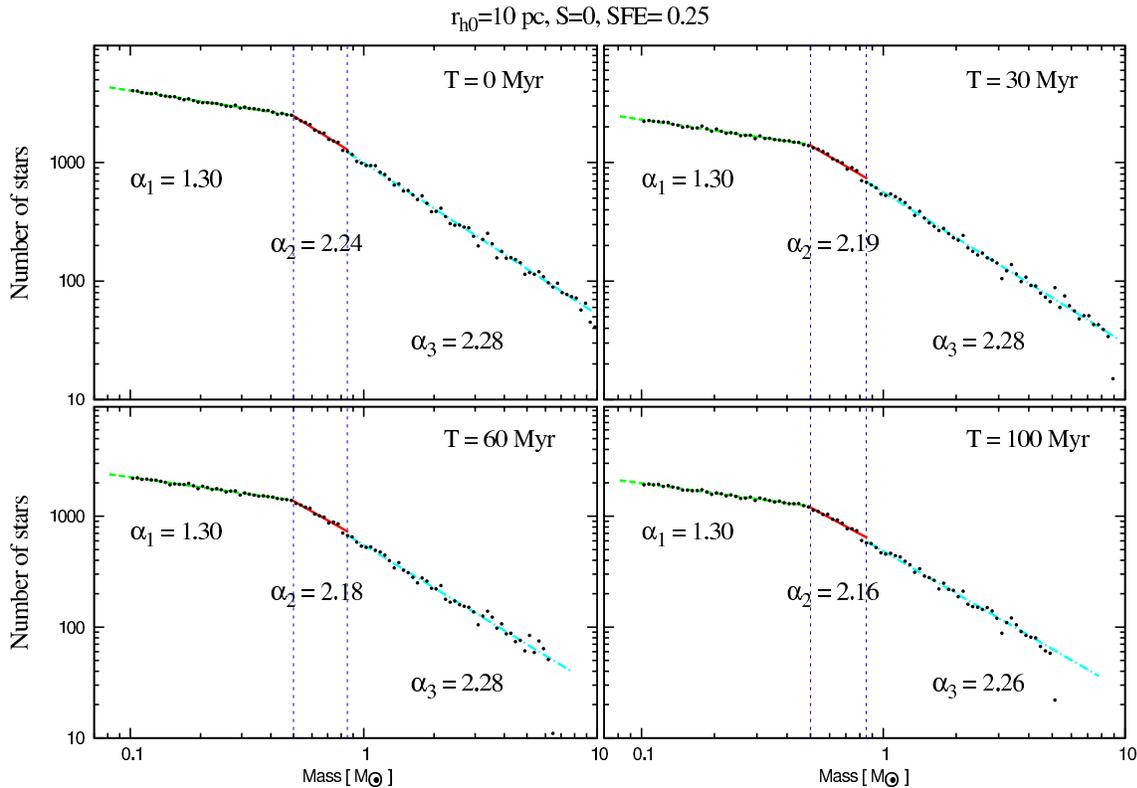}
\caption{Evolution of the global stellar mass function (for main-sequence stars)  inside the tidal radius is computed at each snapshot for a tidally filling cluster without primordial mass segregation. The star formation efficiency is set to $\varepsilon=25\%$. The dashed vertical lines show the positions of $m= 0.5 M_\odot$ and $m= 0.85 M_\odot$. The values of the best-fitting slopes in the intervals, $m \leq 0.5 M_\odot$ ($\alpha_1$),  $0.5 M_\odot \leq m \leq 0.85 M_\odot$ ($\alpha_2$), and for $m\geq 0.85 M_\odot$  ($\alpha_3$) are written in the plots. As can be seen, the effect of gas expulsion on the mass function is negligible for initially not mass-segregated clusters. }
\label{mf_S0}
\end{center}
\end{figure*}

\section{Description of the models}

\begin{figure*}
\begin{center}
\includegraphics[width=170mm]{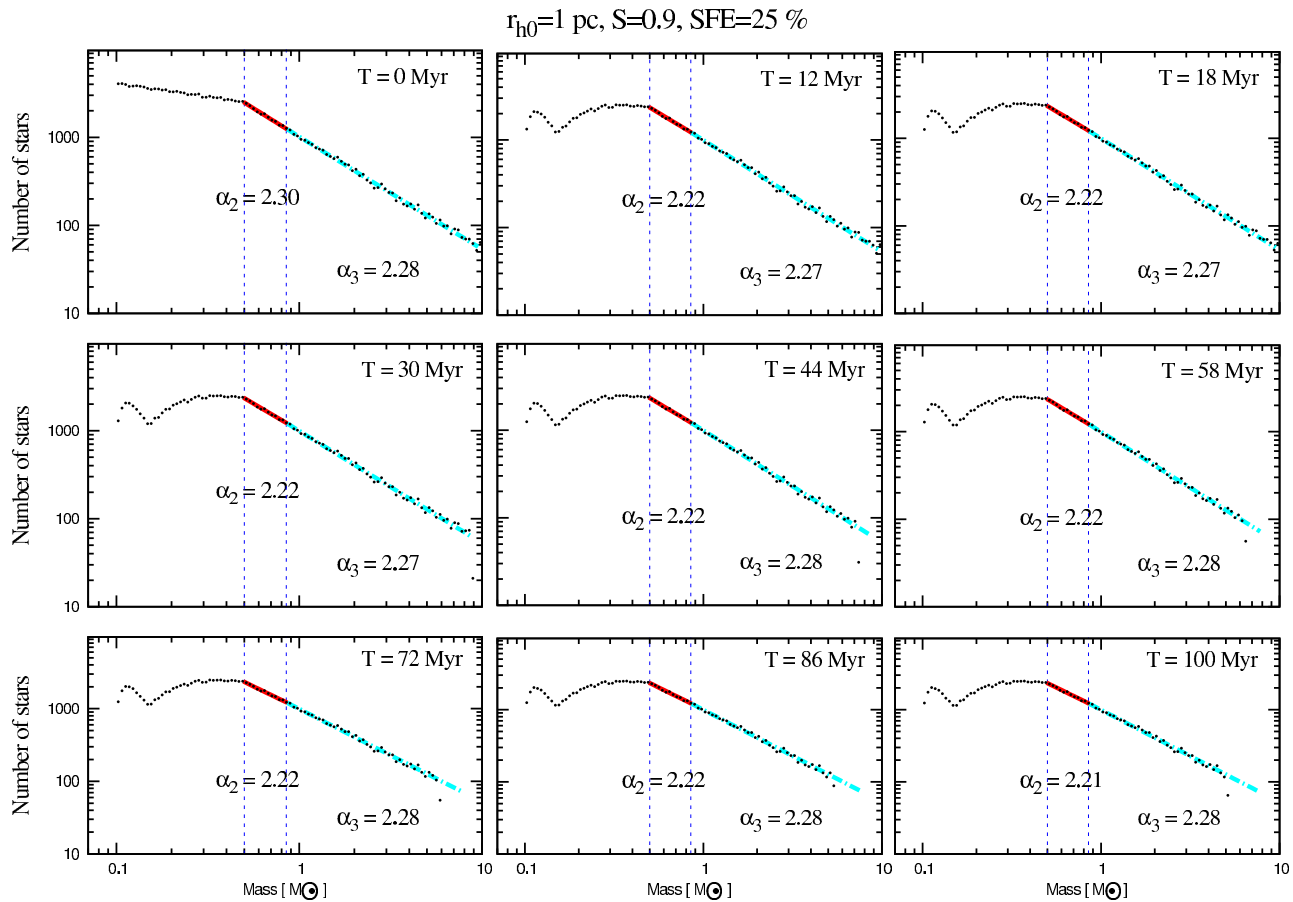}
\caption{Evolution of the global stellar mass function within the tidal radius of an initially tidally underfilling cluster with primordial mass segregation (S=0.9).  The star formation efficiency is set to $\varepsilon=25\%$. The initial half-mass radius of the modeled cluster is $r_{h0}=1$ pc. The dashed vertical lines shows the position of $m= 0.50 M_\odot$ and $m= 0.85 M_\odot$. The slope of the best fit line in this interval, $\alpha_2$ and for $m\geq 0.85 M_\odot$,  $\alpha_3$, are written in the plots. As can be seen the slopes in these mass ranges do not change away from the initial form because the high-mass stars which are located in the inner part of such a compact cluster are less affected by the tidal interaction. }
\label{mf_S9rh1}
\end{center}
\end{figure*}

\begin{figure*}
\begin{center}
\includegraphics[width=170mm]{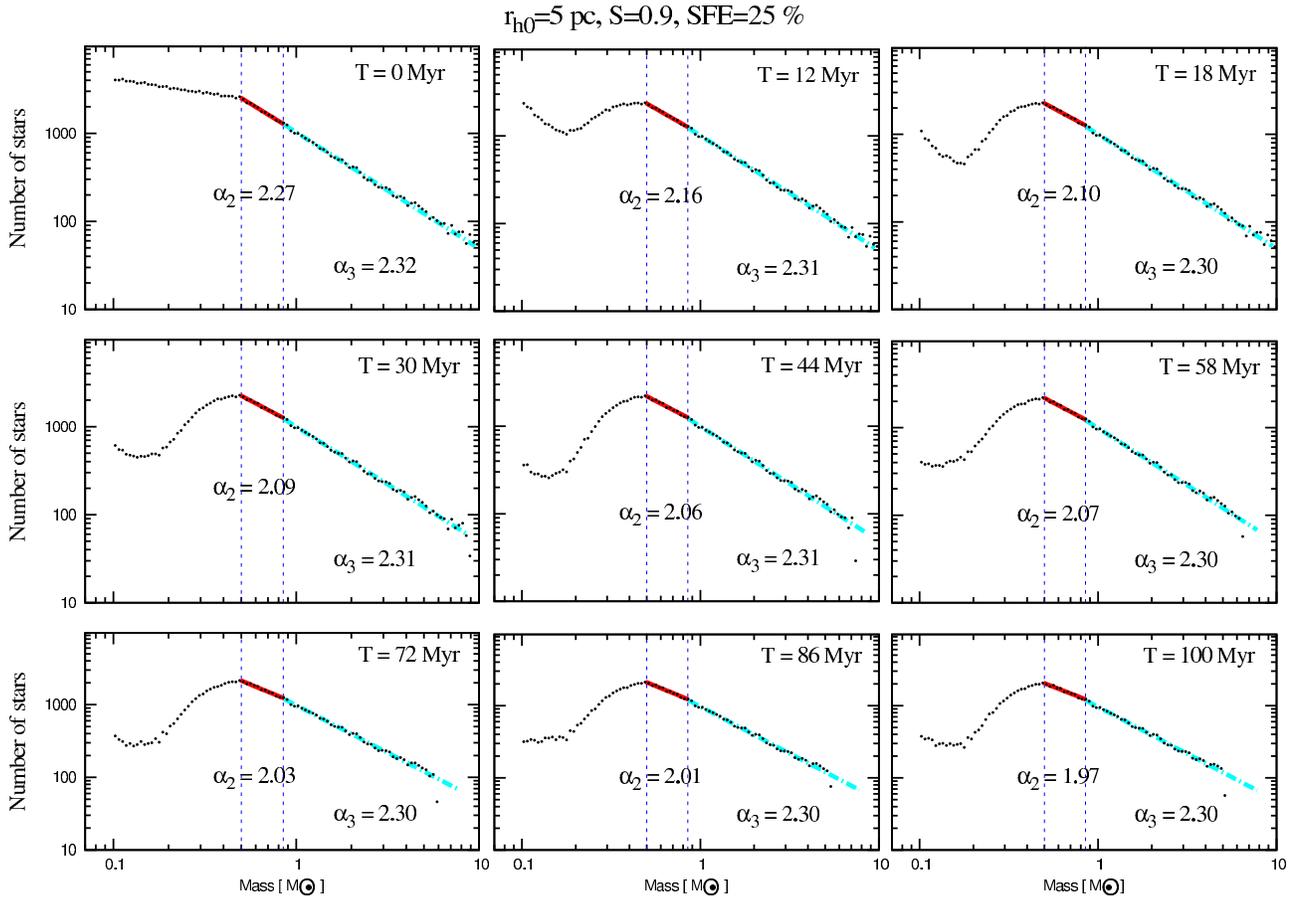}
\caption{The evolution of the global stellar mass function within the time evolving tidal radius of an initially  \emph{marginally tidally underfilling} cluster with primordial mass segregation (S=0.9).  The star formation efficiency is set to $\varepsilon=25\%$. The initial half-mass radius of the modeled cluster is $r_{h0}=5$ pc ($r_h/r_t$=0.09). The dashed vertical lines show the position of $m= 0.5 M_\odot$ and $m= 0.85 M_\odot$. The slope of the best fit line in this interval, $\alpha_2$ (that changes from $\alpha_2=2.27$ at T=0 to $\alpha_2=1.97$ at T=100 Myr), and for $m\geq 0.8 M_\odot$,  $\alpha_3$, are written in the plots. }
\label{mf_S9rh5}
\end{center}
\end{figure*}

\begin{figure*}
\begin{center}
\includegraphics[width=170mm]{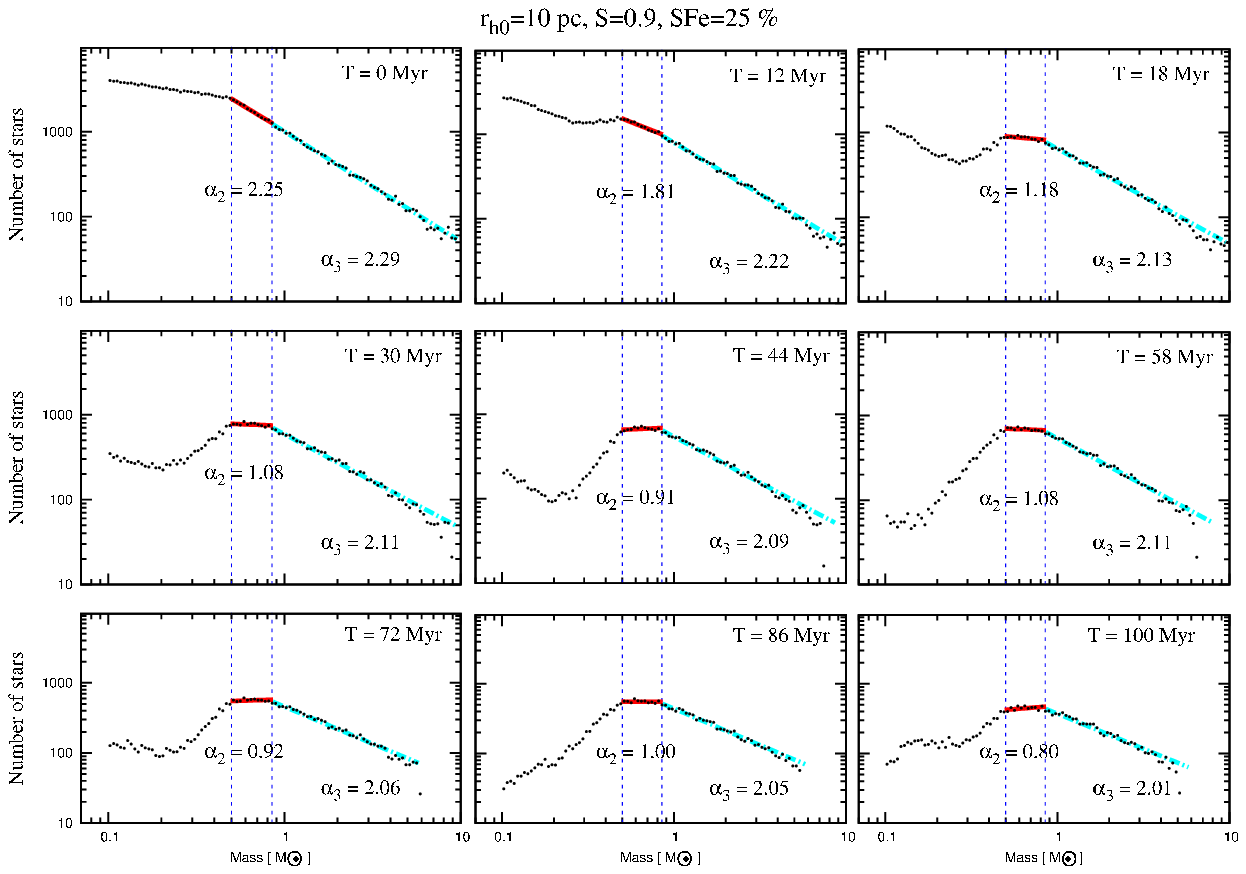}
\caption{The same as Fig. \ref{mf_S9rh5} but for a tidally-filling cluster with $r_{h0}=$10 pc. The slope of the MF in the mass range $m= 0.50 M_\odot$ and $m= 0.85 M_\odot$ significantly changes away from the canonical value of 2.30. The MF shows a nonlinear trend for $m\leq 0.30 M_\odot$, below which it becomes much shallower than the canonical IMF. An explanation for this specific form is represented in Sec. 5  }
\label{mf_S9rh10}
\end{center}
\end{figure*}

The effective dynamical influence of gas-removal from embedded clusters is included as modeling the gas as an additive time-varying potential and accordingly modifying the equations of motion of the stars \citep{kroupa01, Baumgardt07}. I.e., the gas was not simulated directly, instead we mimic the essential dynamical effects of the gas-expulsion process by applying a diluting, spherically-symmetric external gravitational potential to a model cluster. Specifically, we use the potential of the spherically-symmetric, time-varying mass distribution as described by Eq. 2.  Comparing computations treating the gas with the SPH method,  Geyer \& Burkert (2001) have demonstrated that such an analytical approach is physically realistic.

The models presented here assume the pre-gas expulsion stellar plus gas configuration of the cluster models to be in virial equilibrium. This is a simplifying assumption which is necessary to make the models computable, because treating the gas component with a magneto-hydrodynamical (MHD) and radiation-transfer code including stellar feedback (i.e., radiation and matter out flows) to the star forming gas is computationally impossible even for a small cluster of mass larger than about $100\,M_\odot$ (Klessen et al. 1998; Bate \& Bonnell 2004; Bate 2009; Girichidis et al. 2011, 2012; Bate 2012). In reality the earliest phases of embedded cluster assembly are given by the simultaneous inward flow of gas, driven by the deepening potential well of the embryonic embedded cluster, and by the nearly-simultaneous outflow of gas driven by the increasing stellar feedback and magnetic fields generated in the proto-stellar accretion disks.

We use the state-of-the-art collisional $N$-body integration code \textsc{NBODY6} (Aarseth 2003) on GPU computers to perform direct N-body simulations. \textsc{NBODY6} uses a 4th-order Hermite integration scheme and an  individual timestep algorithm to follow the orbits of cluster  members and invokes regularization schemes to deal with the  internal evolution of small-N subsystems.  \textsc{NBODY6}  includes the tidal effects of the Galaxy,  and stellar and binary evolution using the dynamical integrated  SSE/BSE routines developed by Hurley et al. (2000, 2002). It includes the time-varying gas potential to model the gas-expulsion process. The initial condition of all clusters are set up using the publicly available code \textsc{McLuster}\footnote{
https://github.com/ahwkuepper/mcluster.git}(K\"upper et al. 2011).

In this work, all models are located  at a Galactocentric distance of $R_G = 8.5$\,kpc orbiting on a circular orbit with circular velocity of $V_G=220$ km/s. The effect of the external tidal field is adopted to be the same as the Milky Way potential including bulge, disk and a Galactic NFW halo potential (Navarro et al. 1996, 1997) which can be viewed as a rough proxy for the log-normal isothermal phantom dark matter halo expected in Milgromian dynamics (Famaey \& McGaugh 2012; Kroupa 2015). We calculate several models with different initial cluster half-mass radius ($r_h=$1, 5, 7 and 10 pc) to cover two different regimes: tidally-filling ($r_h/r_t\geq0.1$) and tidally-underfilling ($r_h/r_t\leq0.1$) clusters. The $r_h\approx$1 pc clusters are consistent with the radius - mass relation for embedded clusters inferred by \citet{Marks12}. We emphasize that the adoption of the galactic model here merely serves the purpose of defining a computable model.  In reality, the Galaxy would not have existed when the GCs formed and the proto-Galactic tidal field would have been evolving locally rapidly (Marks \& Kroupa 2010).

The number of stars and initial cluster mass in all computed models are set to N=170000 and $M=10^5 M_{\odot}$, respectively. The initial mass function of stars is adopted to be a canonical Kroupa IMF which consists of two power laws with slope $\alpha$=1.3 for stars with masses, $m$, between 0.08 and $0.5 M_{\odot}$ and a slope $\alpha$=2.3 for more massive stars. Stellar evolution is modelled according to the routines by Hurley et al. (2000). The range of stellar masses was chosen to be from 0.08 to 100$M_{\odot}$ and the metallicity is Z$ = 0.001$\,dex (which is a typical value for the young halo GCs).  The evolution time for all models is 100 Myr.  Frank et al. (2012) estimate an age of $11\pm 1$\,Gyr for Pal~4 (as an example of the typical object we study) by adopting [Fe/H]$ = -1.41\pm 0.17$\,dex for the
metallicity from the best-fitting isochrone of \citet{Dotter08}.

For initially segregated models, the mass segregation was set up by using the segregation routine described in \citet{Baumgardt08a}. The degree of segregation (parameterized by $S$) can be chosen to be between 0 and 1, where $S=0$ means no segregation and $S=1$ refers to full segregation, i.e.~ the most massive star sitting in the lowest energy orbit and the second-most massive stars sitting in the second-lowest  energy orbit, and so on. This routine allows to maintain the desired density profile when increasing the degree of mass segregation while also making sure that the cluster is in virial equilibrium. The primordial segregation for all models are set to S=0.9, except one model  which is set to S=0 for comparison. We do not have primordial binaries in our simulated clusters, although binaries created via three-body interactions are automatically included.  Including a significant fraction of primordial binaries is computationally impossible for such massive clusters as moddeled here (see Leigh et al 2015 for details).

Concerning the assumption of primordial mass segregation, recently, Parker et al. (2015) used the final state of SPH simulations of star formation which include feedback from photoionization and stellar winds as the initial condition of their N-body simulations and found that after 10 Myr of evolution,  primordial mass segregation is less common in the simulations when star formation occurs under the influence of feedback.  They found that the ten most massive stars are mass segregated, rather than full mass segregation of the whole cluster.   These simulations concern low-mass clusters with very simplified treatment of feedback, such that it is unclear how indicative they are.  For example,  the outflows launched even by low-mass stars appear to be destructive on a cloud scale. Observations of some young clusters such as ONC (Allison et al 2009), NGC3603 (Pang et al 2013) and Trumpler 14 (Sana 2010)  have shown these clusters to be mass segregated down to around the 4$^{th}$, 20$^{th}$, and 6$^{th}$ most massive stars, respectively, rather than primordial mass segregation of the whole cluster. Thus, while the complete mass segregation in young clusters has not been observed yet,  in this paper, we aim to address possible initial conditions of star-burst clusters in a totally different mass regime a Hubble time ago in low-metallicity gas.

In our calculations, we assume that the SFE does not depend on the position inside the cluster, so both stellar and gaseous components follow the same density distribution initially, which is given by a Plummer model with similar radius. Using a Plummer model is a reasonable physical choice because it is the simplest solution of the collisionless Boltzman equation and also adequately describes the simplest stellar-dynamical objects, globular clusters, known to exist (Plummer 1911)   We assume different  SFEs: $\varepsilon: 25\%, 33\%$. The gas-expulsion delay times for all models are set to $\tau_{D}=0.6$ Myr. The details of the models are given in Table 1.

\section{Results}\label{best-fit}

In this section we present the results from our numerical simulations. The mass-loss in the beginning of the evolution is dominated by gas expulsion and early mass-loss associated with stellar evolution of
massive stars, such that clusters lose about 30\% of their mass within the first 100 Myr by early stellar evolution. Moreover, in the framework of our model, the evolution of a star cluster is affected by several parameters related to the details of the gas expulsion process (like e.g., star formation efficiency, $\varepsilon$ , and gas expulsion time scale, $\tau_{GE}$ ) as well as the strength of the external tidal field, quantified by the ratio of the initial half-mass radius of a star cluster to its tidal radius, $r_h/r_t$.
In addition, the presence of primordial mass segregation significantly affects the global dynamical evolution of star clusters. In tidally limited clusters, primordial mass segregation leads to a stronger expansion, and hence a larger flow of mass over the tidal boundary \citep{Banerjee14}. It may therefore help to dissolve them more rapidly. Tidally-underfilling clusters, however, can survive this early expansion (Vesperini et al. 2009).  The degree of primordial mass segregation and the strength of the tidal field of the host galaxy are therefore  a crucial parameter in the early evolution of GCs. We therefore  compare the evolution of the MF of the clusters by changing the strength of the external field and degree of initial mass segregation.  Table 1 gives an overview of the simulations performed.

\subsection{The influence of residual-gas expulsion on the MF of a primordially not mass-segregated cluster}

We first start with one model without primordial segregation ('rh10-S0-sfe25'). In order to see the maximum effect of residual-gas expulsion, we set the initial half-mass radius of the modeled cluster to $r_h=10$ pc, and therefore it is a tidally-filling cluster ($r_h/r_t=0.18$). We also set $\varepsilon= 25\%$, because the lower the SFE, the stronger is the effect of gas expulsion (GE).  We plot the mass function at different snapshots to see how it evolves with time in the intermediate-mass ($0.50  \leq \frac{m}{M_\odot} \leq 0.85 $) and high-mass  ($m\geq 0.85 M_\odot$) part evolve  with time (Fig.1). We also fitted the MF of main-sequence stars with masses in the three different mass ranges of  $m \leq 0.50 M_\odot$  ($\alpha_1$), $0.50  \leq \frac{m}{M_\odot} \leq 0.85 $  ($\alpha_2$), and $m\geq 0.85 M_\odot$ ($\alpha_3$) with a power law function.

As can be seen, the mass function of the initially non-segregated cluster does not change with time. Just a very slight decrease of the slope due to two-body relaxation driven evaporation of stars occurs in the mass range $0.50 \leq \frac{m}{M_\odot} \leq 0.85 $. This implies that the shape of the MF stays more or less the same, i.e. stars of all stellar masses are equally affected by residual gas-expulsion.
In Figure \ref{alpha} (see model 'rh10-S0-sfe25') we also show the time evolution of $\alpha_2$. It can be seen that the MF flattening occurs on a much longer timescale for the non-mass-segregated system.

\subsection{The influence of residual-gas expulsion on the MF of primordially mass-segregated clusters}

We assigned a quite extreme degree of segregation of $S=0.9$ for all primordially segregated models, in order to show the maximum influence mass segregation can realistically have.

\subsubsection{Tidally-underfilling clusters}

Because of stellar evolution post gas-expulsion clusters experience a further expansion such that clusters with a higher degree of primordial mass segregation experience a larger jump in the half-mass radius within the first 100 Myr of evolution. Almost 30\% of the initial mass is lost owing to stellar evolution on a timescale of 100 Myr.
Fig. \ref{mf_S9rh1} depicts the effect of gas expulsion on the MF of a tidally-underfilling cluster with an initial half-mass radius of $r_{h0}=1$ pc and a filling factor of $r_h/r_t = 0.02$. As can be seen, the MF in the range  $m \geq 0.5 M_\odot$ still resembles the IMF and does not show a strong change of the slope, while in the mass range  $m \leq 0.5 M_\odot$ there is a significant change in the slope.

In fact, if the tidal field is weak (i.e. $r_h/r_t = 0.02$, see Table 1), even a model with a SFE as low as 25\% keeps the majority of its mass and undergos very little mass loss due to the tidal interaction. According to the final mass demonstrated in Table 1, a very small fraction of mass is lost due to dynamical evolution caused by the gas expulsion process. In other words, the initially tidally-underfilling models experience a weak tidal field, and hence keep most of their stars and retain their IMF slope in the mass range $m \geq 0.5 M_\odot$.

Fig. \ref{mf_S9rh5} shows different snapshots of the mass function of a \emph{marginally tidally-underfilling} cluster ($r_h/r_t = 0.09$ with $r_{h0}=5$ pc) with a primordial mass segregation with S=0.9 and a star formation efficiency of $\varepsilon= 25\%$. As can be seen, gas expulsion significantly changes the slope of the mass function only in the low-mass part ($m \leq 0.50 M_\odot$). The mass function at the high-mass end does not change, while the slope in the intermediate mass range, $0.50 M_\odot \leq m \leq 0.85 M_\odot$, changes from $\alpha_2=$2.3 to 2. Note that this model is not an extreme case of tidally-underfilling systems, but it is an intermediate case between the two regimes of tidally-filling and tidally-underfilling clusters. Figure \ref{alpha} shows the time evolution of $\alpha_2$ for this model.

\subsubsection{Tidally-filling cluster}

For clusters initially tidally-filling their Roche lobes (i.e. models with $r_h/r_t\geq 0.1$) and primordially mass segregated, a substantial fraction of the initially bound mass is lost simply due to stellar evolution and  expansion of the cluster driven by gas expulsion. Our simulations show that a given amount of mass loss leads to a stronger MF flattening than for tidally-underfiling clusters.  The strong tidal fields lift even the clusters with a large SFE away from the initial IMF slope.

Fig. \ref{mf_S9rh10} depicts the evolution of the slope of the mass function over the whole range of stellar mass at different snapshots for a tidally-filling, primordially mass-segregated cluster with $r_{h0}=10$ pc, i.e. model 'rh10-S0.9-sfe25'. The slope was determined from a fit to the distribution of stars in two different mass ranges, $0.50 M_\odot \leq m \leq 0.85 M_\odot$ ($\alpha_2$), and $m\geq 0.85 M_\odot$ ($\alpha_3$). Fig. \ref{mf_S9rh10} confirms that the slope of the mass function for tidally-filling clusters changes in the intermediate-mass range as well as in the low mass part, by getting shallower. Figure \ref{alpha} shows the evolution of the MF-slope in the intermediate-mass range for all computed models.

Comparing the last two models in Table 1 and Fig. \ref{alpha}, it can be seen that for initially segregated clusters, as the SFE decreases, the slope of the MF in the low-mass part decreases. This is easy to understand: The lower the SFE gets, the stronger is the dynamical effect of the gas expulsion process.

The oscillatory trend of $\alpha_2$ in Fig. \ref{alpha}, particularly notable for the mass-segregated tidally-filling clusters, are  related to the oscillatory trend of the mean velocity dispersion (Fig. \ref{v_S9rh10}) of the outer layers which are dominated by the low-mass population of stars (see Sec. 5 for more details). According to Fig. \ref{massloss}, after an early expansion driven by gas expulsion, the half-mass radius increases by a factor of 2 (i.e., $r_h=$20 pc), and the total mass within the tidal radius decreases to $M=20000 M_\odot$ at $T=100$ Myr, so the dynamical time scale of the cluster is equal to $t_{cross}\simeq30$ Myr. This value is nearly equal to the period of the oscillations as can be seen in Figs. \ref{alpha}, \ref{v_S9rh10} and \ref{vr_S9rh10}.

\subsection{Comparison with observation: $\alpha-c$ plane }

\begin{figure}
\begin{center}
\includegraphics[width=85mm]{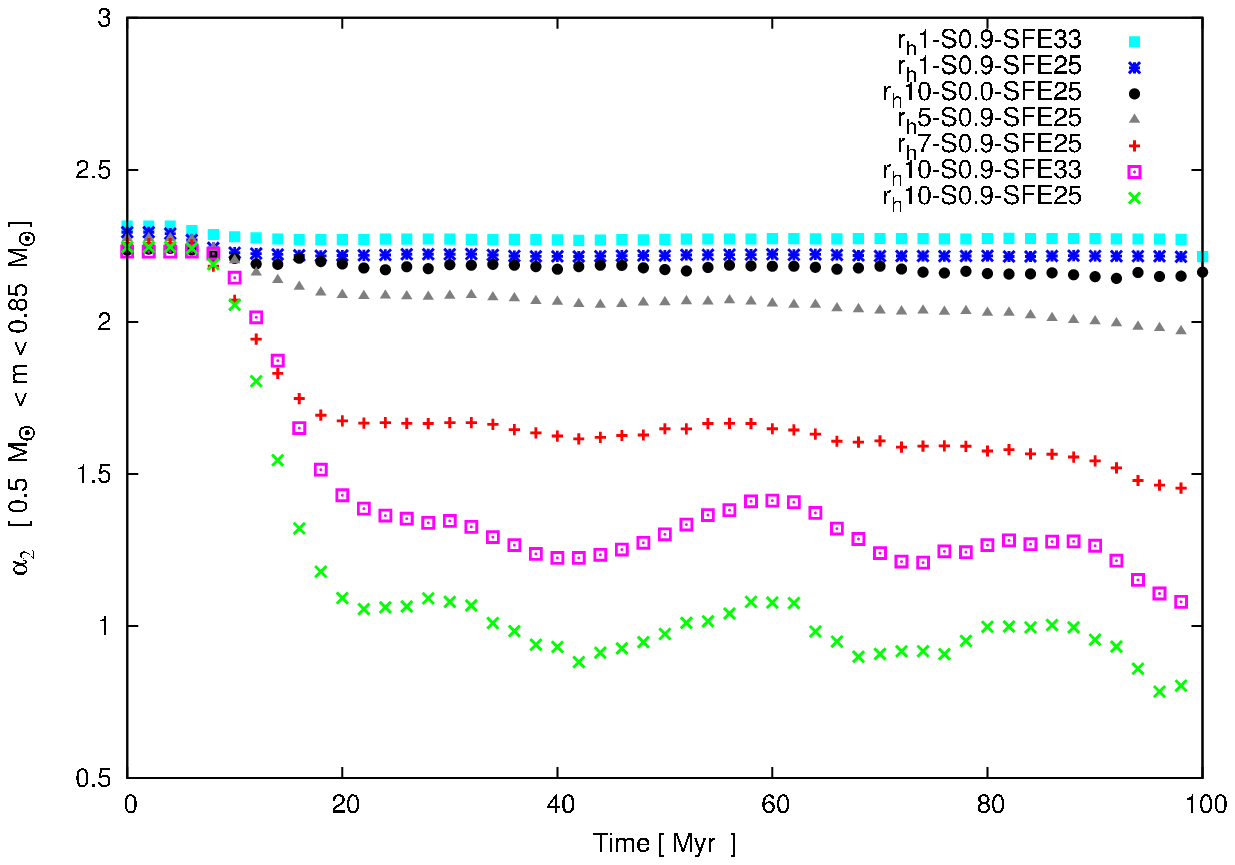}
\caption{The time evolution of the MF-slope, $\alpha_2$, for main sequence stars within the tidal radius in the mass range $0.50 M_\odot \leq M \leq 0.85 M_\odot$. The initial number of stars in all models is N=170000 (corresponding to $M=10^5 M_\odot$) with an input Kroupa canonical IMF with the slope of $\alpha_2=2.3$. The initial half-mass radius, star formation efficiency, and degree of primordial segregation (i.e., $ S=0$ and $ S=0.9$) are indicated in the model name in the legend.  As can be seen only tidally filling clusters with primordial mass segregation show a significant flattening in the slope of the MF.  In the  primordial mass segregated models, the lower SFEs lead to a shallower MF-slope.}
\label{alpha}
\end{center}
\end{figure}

\begin{figure}
\begin{center}
\includegraphics[width=85mm]{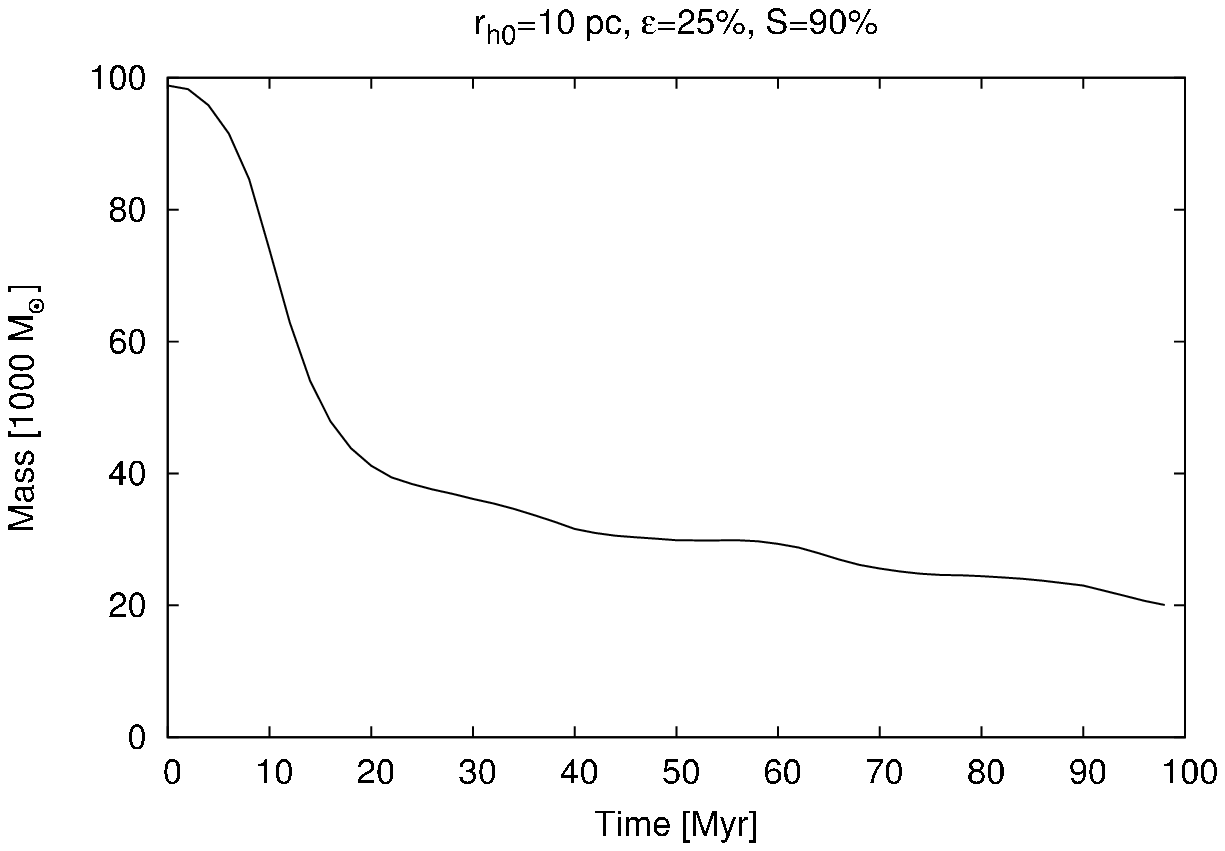}
\includegraphics[width=85mm]{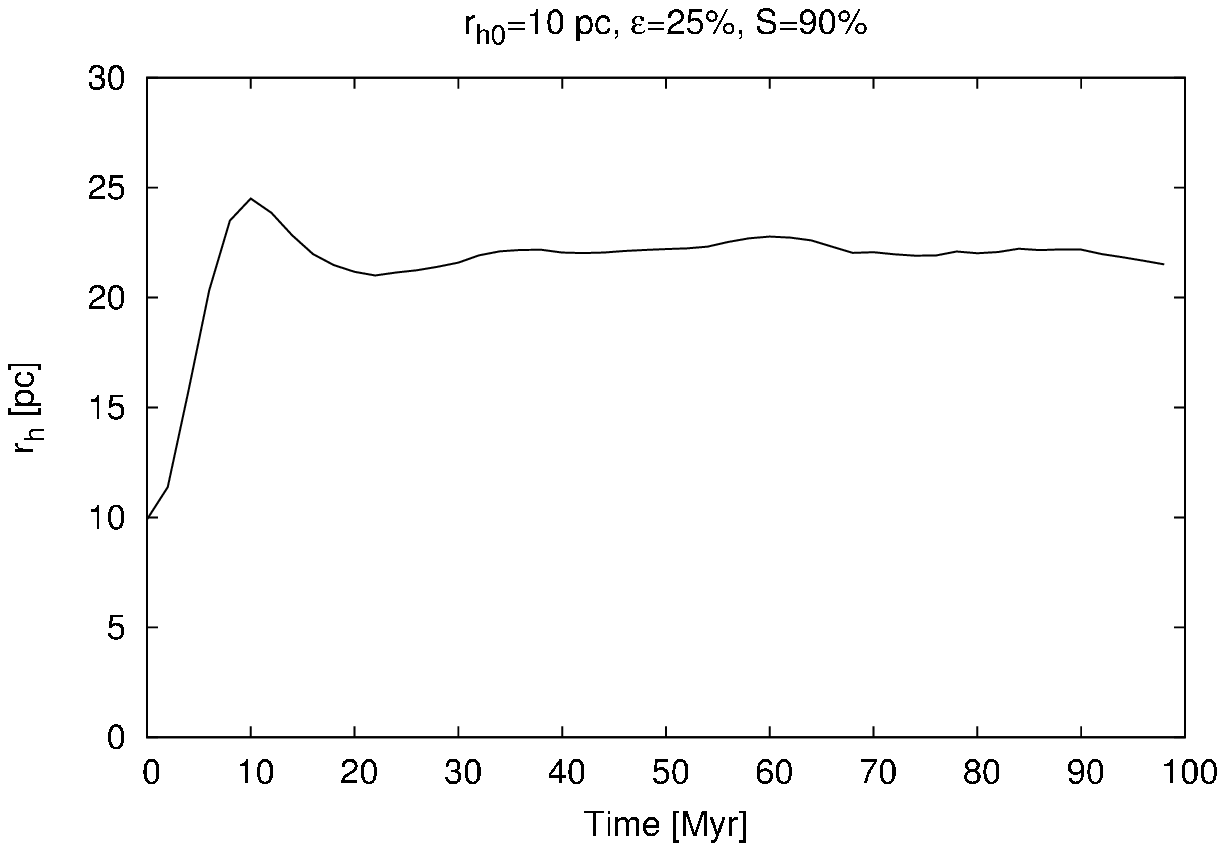}
\caption{The evolution of total mass (upper panel) and half-mass radius (lower panel) with time for simulated star cluster "rh10-S0.9-sfe25".  The cluster loses about 80\% of its mass by early rapid gas expulsion and stellar evolution.}
\label{massloss}
\end{center}
\end{figure}

\begin{figure}
\begin{center}
\includegraphics[width=85mm]{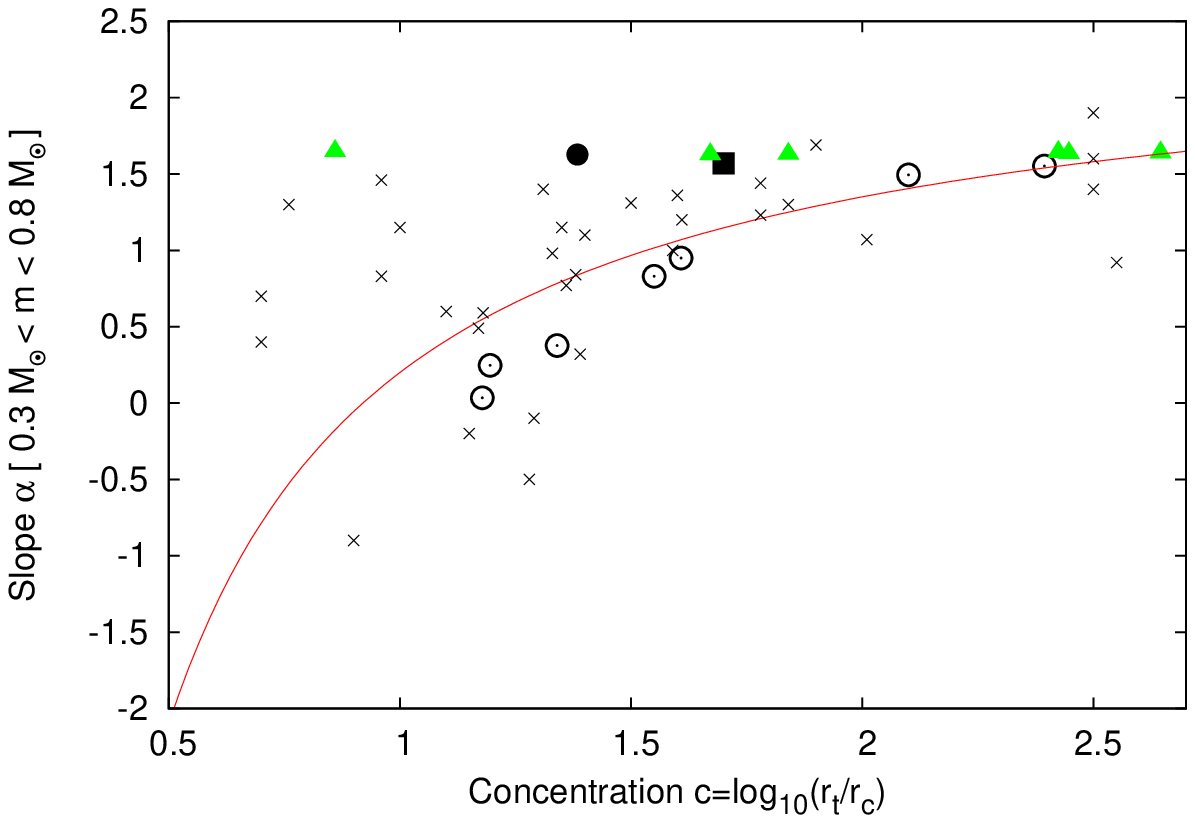}
\caption{The slope $\alpha$ of the MF at the end of the N-body integrations ($T=100$ Myr) in the mass-range 0.3 - 0.8 $M_\odot$  as a function of the cluster concentration, $c = \log_{10}(r_t/r_c)$, for all calculated models listed in Table 1 (circles with central dots).  The solid line is an eye-ball fit to the observed data (crosses) from De Marchi, Paresce \& Pulone (2007) and \citet{Paust10}. The filled square marks the primordially segregated modeled cluster \emph{without gas expulsion},"rh10-S0.9-NoGE", the filled circle represents the non-segregated model "rh10-S0-sfe25", and the filled triangles represent simulations without gas expulsion and primordial mass segregation. }
\label{alpha-c}
\end{center}
\end{figure}

Fig. \ref{alpha-c} compares the influence of gas expulsion on the MFs of computed models, with the eye-ball fit to the observed data from De Marchi, Paresce \& Pulone (2007, hereinafter MPP ) with the functional form of $\alpha(c)=\frac{-2.3}{c}+2.5$, supplemented with additional global MF slopes taken from Paust et al. (2010). The canonical Kroupa IMF has $\alpha=1.72$ over this mass-range ($0.3 - 0.8 M_\odot$).  \emph{The models with gas expulsion agree well with the general trend that less concentrated clusters typically show a shallower MF.} The non-mass segregated model (filled circle), and a test model without gas expulsion (filled square), are not able to reproduce the observed trend confirming the conclusion by Leigh et al. (2013). These results confirm, by direct N-body simulations, the conclusions reached by \cite{Marks08} based on simplified models. In order to justify our conclusion we calculated five additional simulations without gas expulsion and primordial mass segregation for different values of the concentration parameters in the range of $c=0.85 - 2.65$. The results are demonstrated as  triangles in Fig. \ref{alpha-c}. As can be seen these models keep their initial MF-slope and independent of the value of the concentration parameters all lie above the observed data points.


From Fig. \ref{alpha}, we see that the change in $\alpha_2$ over time is small/negligible for tidally under-filled clusters and is more noticeable for tidally marginally-filled and filled clusters. Note that the declining and oscillating trend of $\alpha_2$ for the tidally filling clusters in Fig. \ref{alpha} would continue well beyond 100 Myr and will \emph{moderately} change their positions in the $c-\alpha$ diagram, they will still be consistent with the MPP trend since the latter has some scatter anyway. This is because the estimated two-body relaxation times for all models with an altered MF are larger than one-third of the Hubble time at $100$ Myr. Therefore, the $\alpha_2$ values will likely decline further due to unbound escaping stars over the intended mass range still remaining within $r_t$.

\subsection{Implications for the very young MW}

Our simulations show that only in the case where a cluster ends up with a very  large $r_h/r_t$, will the low mass stars leave the cluster. Note that all clusters in Table 1 which have a significantly altered MF due to gas expulsion end up with final 3D half-mass radii of about $\simeq$21 pc. This is compatible with the typical size of remote halo globular clusters such as Palomar 4 and Palomar 14, but they are still larger than the size of the  clusters in the MPP  plot with a median half-mass radius around $<r_h>=8.9 \pm 5.9$ pc.

Although the present-day radii of the MPP clusters do not fit a tidally-filling scenario, this does not mean that the particular clusters that have a damaged MF were not tidally filling during the fist 10-100 Myr of their life when the expulsion of residual gas caused violent relaxation and expansion. The present-day population II halo GCs which formed with a higher metallicity, and thus typically after the GCs with a lower metallicity, most probably formed in an environment which had larger fluctuations in the tidal field (Marks \& Kroupa 2010, their fig. 12).

Also, during the first few Gyr the MW did not exist in the way we know it today. Observations have revealed that the size and the mass content of galaxies change significantly with redshift such that the sizes of the galaxies at high redshifts are smaller in comparison with galaxies of similar mass in the local universe (e.g., \citealt{Franx08, Williams10, Mosleh11, Law12, Mosleh13}). I.e., the fluctuation in the  pre-MW potential grew with time  as it began assembling during the first few Gyr and it is only during the past 8-10 Gyr that the MW finally settled down in the form we know it today. So, when the GCs were relatively young, their environment would have been much more dramatic. Some, but not all, of these GCs would have formed in a strong local tidal field such that they lost a significant fraction of their stellar population as a result of the expulsion of their residual gas. For example, a very young GC with a mass of $M_{\rm GC}=10^7\,M_\odot$ would have formed with an embedded half-mass radius of about 0.8 pc (assuming the radius-mass relation of Marks \& Kroupa 2012). Expulsion of 67~per cent of unused gas would lead to an expansion to a half-mass radius of about 3~pc, and further expansion through stellar evolution would continue. If the IMF was top-heavy (Marks et al. 2012), then the expansion would be larger still. If this young GC was mass-segregated and immersed in a tidal field from a nearby extremely massive molecular cloud with a mass of, for example, $M_{\rm MC}=10^8\,M_\odot$, which is in the process of forming another GC, then the tidal radius would be $r_t\approx D\,(M_{\rm GC}/M_{\rm MC})^{1/3} = 23\,$pc such that this GC would have had $r_h/r_t>0.13$ if it was located $D=50\,$pc away from the cloud. Such events and configurations would have persisted during the first 0.5-1~Gyr of the evolution of the proto-MW and the above case would exist for 10--100~Myr given that the just-hatched GC would be approximately co-moving with the evolving nearby molecular clouds during half an orbit through the forming proto-MW. Present-day instances of such processes and events are evident in the Antennae galaxies where massive star cluster complexes containing hundreds of massive clusters within region with a diameter of 300 pc are forming \citep{Kroupa98, Wilson00, Wilson03, Bruns11, Johnson15}.

Since the young GCs did not form in a smooth environment, they must have been exposed to strongly fluctuating tidal fields in the very early pre-MW potential.
"R136" is a good example: it is the most massive compact cluster in a whole cluster
complex. There are many lesser clusters that formed in the "R136" star-burst region.  The "Arches" cluster is in fact in a very strong tidal field,
and most probably also did not form alone. However, these cases are almost two or three orders of magnitude lesser in mass scale than what was occurring in the
very young pre-MW. So, the damaged GCs (i.e., the ones with low $c$ and depleted MF) may well have been the lesser clusters in a
star-burst region orders of magnitude more intense than anything in the present-day local Universe.

It should be noted that the correlation is not a pure time-sequence. But the more damaged clusters are typically more metal rich (Marks \& Kroupa 2010). This means they typically formed a few hundred Myr later than the first, metal poorest clusters which are also least damaged and would have formed in
an overall smoother potential (weaker tidal fields) and more compact due to the lower metallicity.

The higher$-[Fe/H]$  clusters are further inside the MW, which means they formed slightly  later and under more violently varying tidal fields, leading to the MPP correlation. Note that this correlation only means that at low c (typically larger [Fe/H]) there is a wider spread of alpha values than at high c (typically lower [Fe/H]). The higher-metallicity clusters typically but not exclusively formed more tidally filling due to larger mass concentrations nearby them, and they formed somewhat deeper in the proto-MW potential.

Other mechanisms may contribute to the MPP trend. For example, by assuming a high retention fraction of stellar-mass BHs after supernovae which prevent the clusters from going into core collapse for nearly a Hubble time,  \cite{Lutzgendorf13} explained  the MPP relation without invoking primordial mass segregation and gas expulsion. They found  that a cluster hosting a high number of stellar-mass black holes  increases the escape rate of high-mass stars from a star cluster, implying that the relative depletion of the mass function at the low-mass end proceeds less rapidly.

Two-body relaxation plus a tidal field, and any process that can modulate the cluster expansion according to concentration (more concentrated clusters expand less) can produce a MPP like trend.  Furthermore,  the observed metallicity trend of the MPP clusters (more damaged GCs have typically higher $[Fe/H]$) needs to be understood in terms of the  stellar-mass BH scenario. In the long term, neutron stars (NSs) will also influence the cluster's expansion. It is to be noted that the BHs and NSs influence mostly the central part of the cluster, the expansion (and contraction later) becomes less prominent as one goes outwards. In any case, BHs (and NSs) can keep the cluster from core collapsing for a long time. Their effect in terms of metallicity can be more subtle. In fact, two opposite trends seem to be there in the $\alpha-[Fe/H]$ correlation:
\begin{itemize}
\item  With lower metalicity, BHs are more massive and would expand the cluster more initially and that tends to produce a trend which is opposite to the MPP's $\alpha-[Fe/H]$ correlation (Marks \& Kroupa 2010).
\item On the other hand, more massive BHs eject each other more efficiently and, in the long run, will be more depleted which is favorable for the metallicity trend.
\end{itemize}
The asymptotic behaviour of stellar remnants in star clusters and their effects are still largely unexplored. However, straightforward canonical stellar dynamical models have been implying that the MPP relation is unexpected (Leigh et al. 2013) unless gas expulsion and mass-segregation at birth with a strongly varying tidal field is taken into account.

\section{Evidence for primordial mass segregation}

As we have shown  in a strong-enough tidal environment, a strongly flattened slope of the stellar MF in the low- ($m\leq 0.5 M_\odot$) and intermediate-mass ($\simeq 0.50-0.85 M_\odot$) regime results after the rapid expansion driven by  residual-gas
removal and massive star evolution.  As can be seen in Figs. \ref{mf_S9rh1}-\ref{mf_S9rh10}  such a flattening is the exclusive feature that occurs only for initially mass-segregated clusters for both regimes, tidally filling and tidally underfilling.  Therefore the imprint of residual gas expulsion, as direct evidence of primordial mass segregation,  might be visible in the present-day MF. Therefore, this can be interpreted as an observational consequence of gas expulsion on the MF in the low mass end confirming the conclusions of \cite{Marks08}.

It should be noted that, apart from the particular shape (the dip near $\simeq 0.2 M_{\odot}$) of the MF in the mass range $m\leq 0.5 M_\odot$,  a cluster which follows the MF shown in, like e.g.,  Fig. 4 ($(r_h/r_t)_0=0.18$, with $S=0.9$) would end up with a mass-to-light ratio smaller by a factor of $0.6$ (due to the depletion of low-mass stars and remnants) compared to the case of Fig 1 ($(r_h/r_t)_0=0.18$, with $S=0.0$).  Of course, this conclusion depends on the way we set up the initially segregated clusters. If we initialize clusters such that there is not such a large degree of mass segregation among the low-mass stars, but only the high mass stars are near the cluster centre, the results will be changed.  In this section we detail our calculations to find an explanation for the particular shape of the mass function in the mass range $m\leq 0.5 M_\odot$.

\begin{figure*}
\begin{center}
\includegraphics[width=85mm]{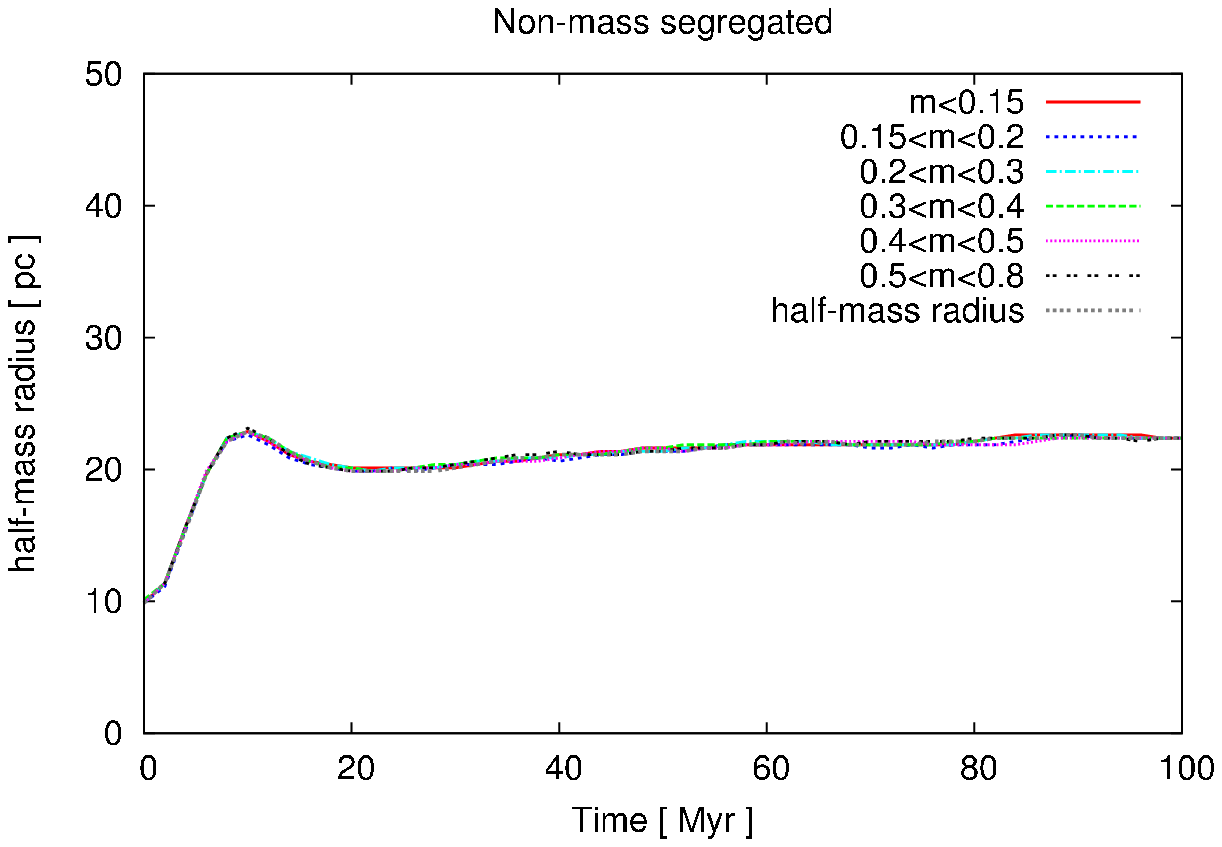}
\includegraphics[width=85mm]{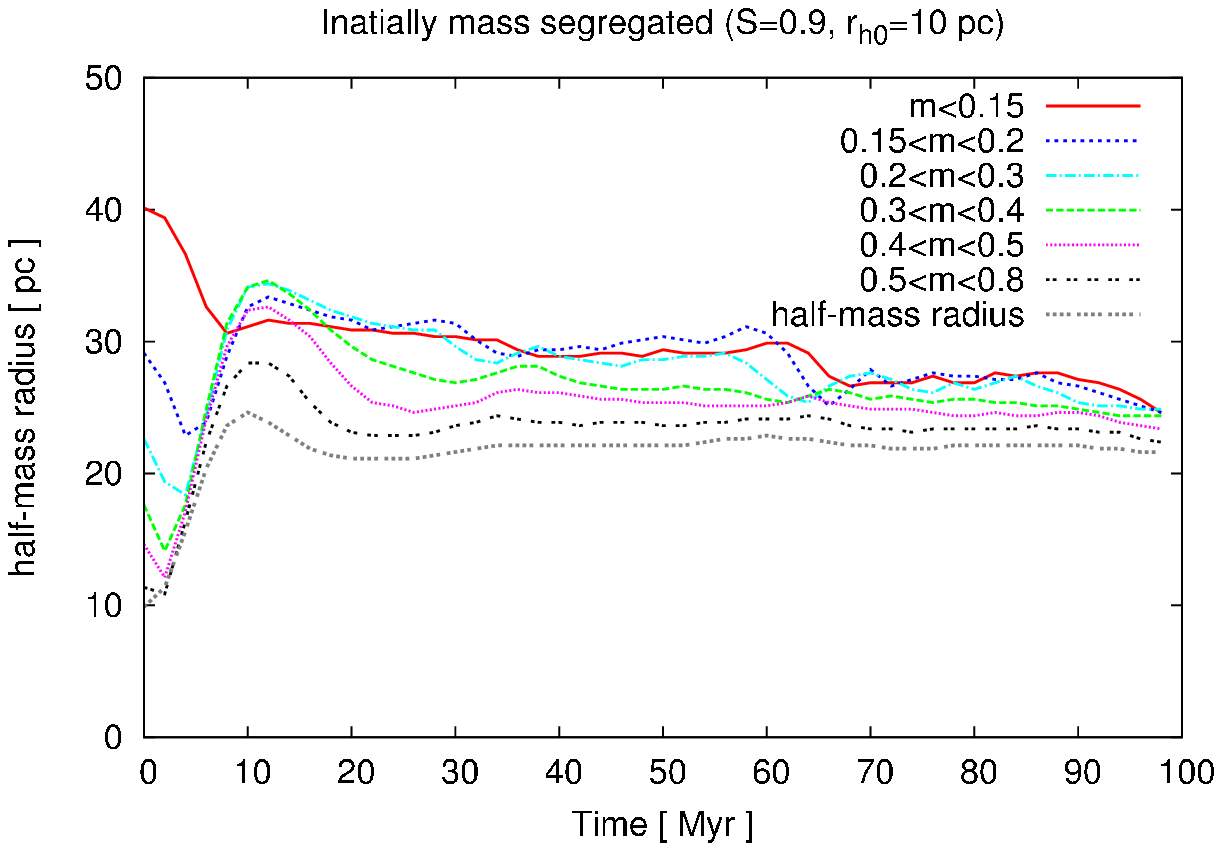}
\caption{   The evolution of the half-mass radii of 6 mass-bins chosen within the range 0.08 to 0.80 $M_\odot$, for two computed tidally-filling models, 'rh10-S0-sfe25' (left: without primordial mass segregation) and 'rh10-S0.9-sfe25' (right: with primordial mass segregation).  The half-mass radius of the cluster is also plotted for comparison. Note that in the non-segregated model, stars with all stellar masses are distributed with the same half-mass radius being equally affected by gas expulsion. In the case of the primordially segregated model, the stars in the mass range $m\leq 0.15 M_\odot$ move inward during the first 10 Myr and are hence less prone to escape from the cluster, while the next inner shells containing the stars with higher mass ($0.15 M_\odot \leq m\leq 0.25 M_\odot$ ) move outwards after an initial brief contraction phase strengthening the possibility of escape from the cluster. This is reflected in the stellar mass function of the cluster (Figs. 3-5) as forming a minimum at $m= 0.2 M_\odot$.  Note that the initial separation of half-mass radii of the different mass-bins is reduced as a result of violent relaxation driven by rapid residual gas-expulsion and all layers tend with increasing time towards the same steady half-mass radius after re-virilization. }
\label{rh_S9rh10}
\end{center}
\end{figure*}

Fig. \ref{rh_S9rh10} (right panel) shows the evolution of the 3D half-mass radius of 6 different mass-bins within the range 0.08 to 0.8 $M_\odot$ for a computed model with primordial mass segregation. For comparison purpose we compute an identical model without primordial mass-segregation (left panel).
Both models are tidally filling with an initial concentration of $(r_h/r_t)_0=0.18$. In the not mass-segregated model, all stars of different mass-bins are equally affected by early gas removal, having the same half-mass radius and they re-virialize in 20 Myr. But, in the case of the primordially mass segregated model, the outer parts which are filled by stars in the mass range $m\leq 0.2 M_\odot$ fall back into the inner regions, while the next inner shells containing the stars with higher masses ($m \geq 0.2 M_\odot $ ) move outwards enhancing the escape rate of stars in this range from the cluster.

Fig. 8 shows why a minimum can be seen in the stellar mass function near $0.2 M_\odot$ with a rise towards lower masses (Figs \ref{mf_S9rh1}-\ref{mf_S9rh10}): In a fully mass segregated and virialised initial cluster the least massive stars ($\leq 0.1 M_\odot$) occupy the least bound orbits at largest radii such that they can only fall towards the cluster centre. They do this despite the expulsion of residual gas. More massive stars are lost from the cluster due to gas expulsion, while the least massive stars are preferentially accumulated. The initial separation of half-mass radii of the different mass-bins vanishes as a result of phase mixing and two-body relaxation and all layers tend to have the same steady half-mass radius long-term after re-virilization.

\begin{figure*}
\begin{center}
\includegraphics[width=85mm]{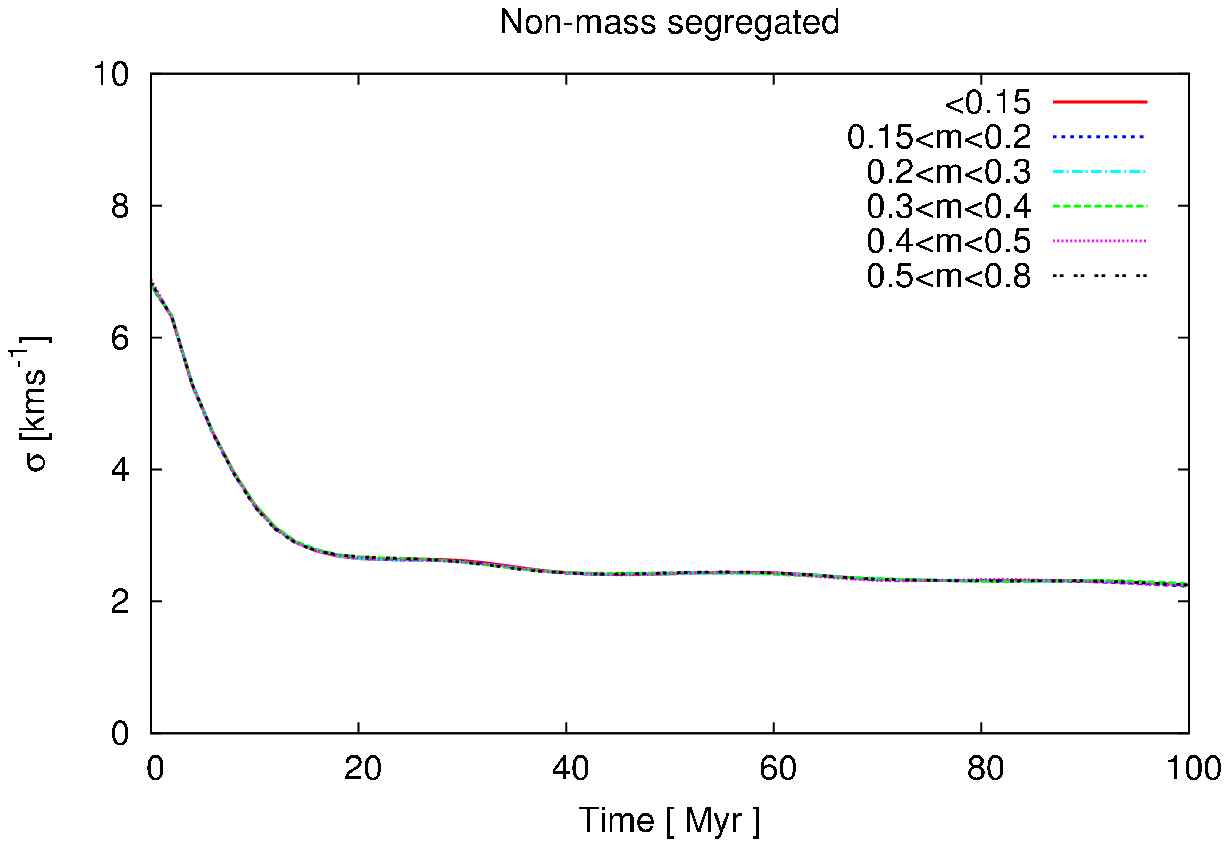}
\includegraphics[width=85mm]{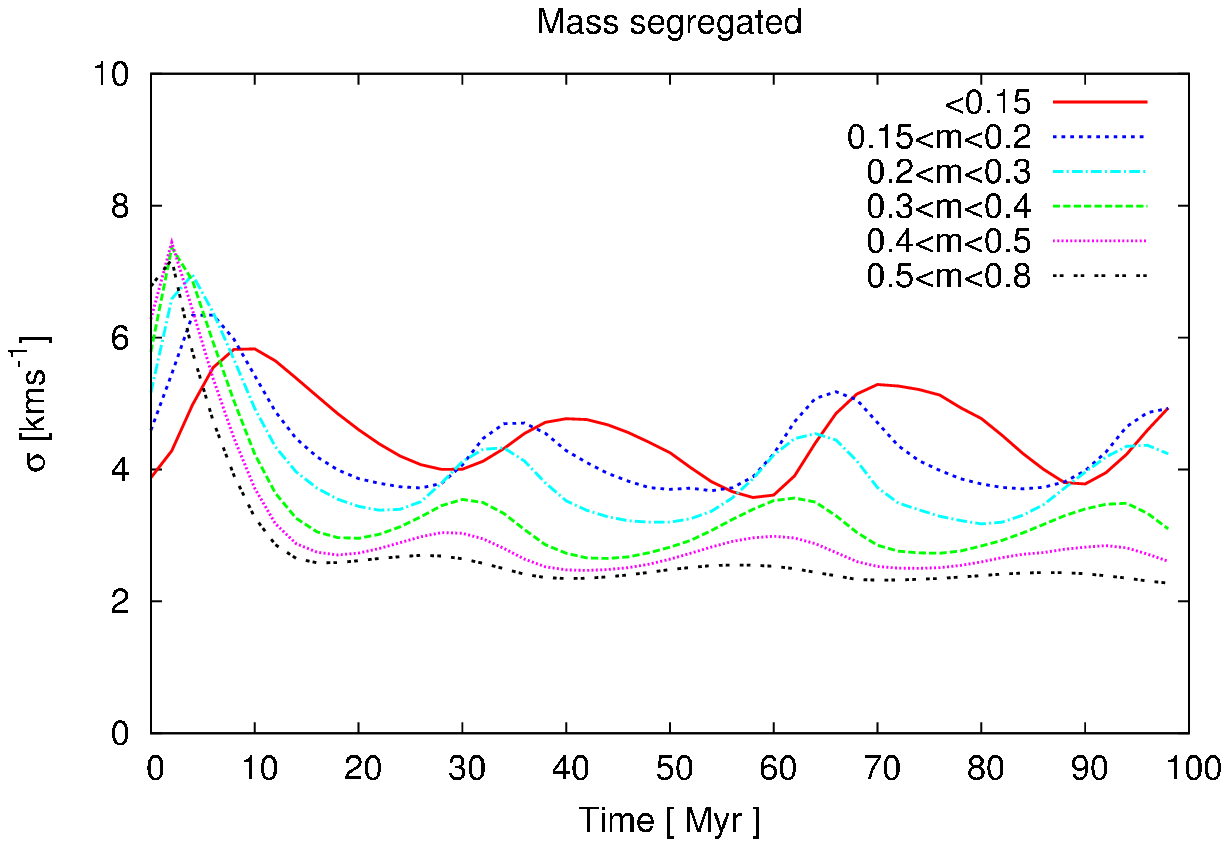}
\caption{Left: The evolution of the velocity dispersion of stars within the tidal radius for different mass-bins as in Fig. \ref{rh_S9rh10}. Left: For a non-segregated model the global velocity dispersion of all stars in all mass-bins equally decreases with time (no difference for different mass-bins) owing to a rapid expansion with timescale $\tau_{GE}$ driven by gas expulsion. After a few crossing times the cluster reaches its steady equilibrium velocity dispersion.  Right: the same as the left panel but for the primordially mass segregated model.  Note that the inner shells (i.e., the higher mass bins) initially have a larger velocity dispersion consistent with the virial equilibrium state. After gas-expulsion and corresponding dilution of the gravitational potential well, owing to the more rapid expansion of the inner shells (with an initially higher velocity dispersion),  the order of the layers is inverted in terms of their velocity dispersion. The oscillation arises from the outer low-mass stars orbiting through the expanded cluster which has a two-body relaxation time of 10 Gyr (Table 1). }
\label{v_S9rh10}
\end{center}
\end{figure*}

Fig. \ref{v_S9rh10} shows the evolution of the mean velocity dispersion within the tidal radius of the 6 mass-bins over 0.08 - 0.80 $M_\odot$ for the same models as shown in Fig. \ref{rh_S9rh10}. In the non-segregated model, the mean velocity dispersion of all stars with different masses equally decreases because of the rapid expansion of the star cluster caused by rapid gas removal.  In the case of the primordially mass segregated model,   and as explained above, the least massive stars begin to fall towards the cluster centre. Their velocity dispersion increases while the more massive stars are partially lost from the clusters. The least massive stars retain their low-energy orbits after re-virialization thereby oscillating through the cluster. Their velocity dispersion therewith oscillates.

\begin{figure*}
\begin{center}
\includegraphics[width=85mm]{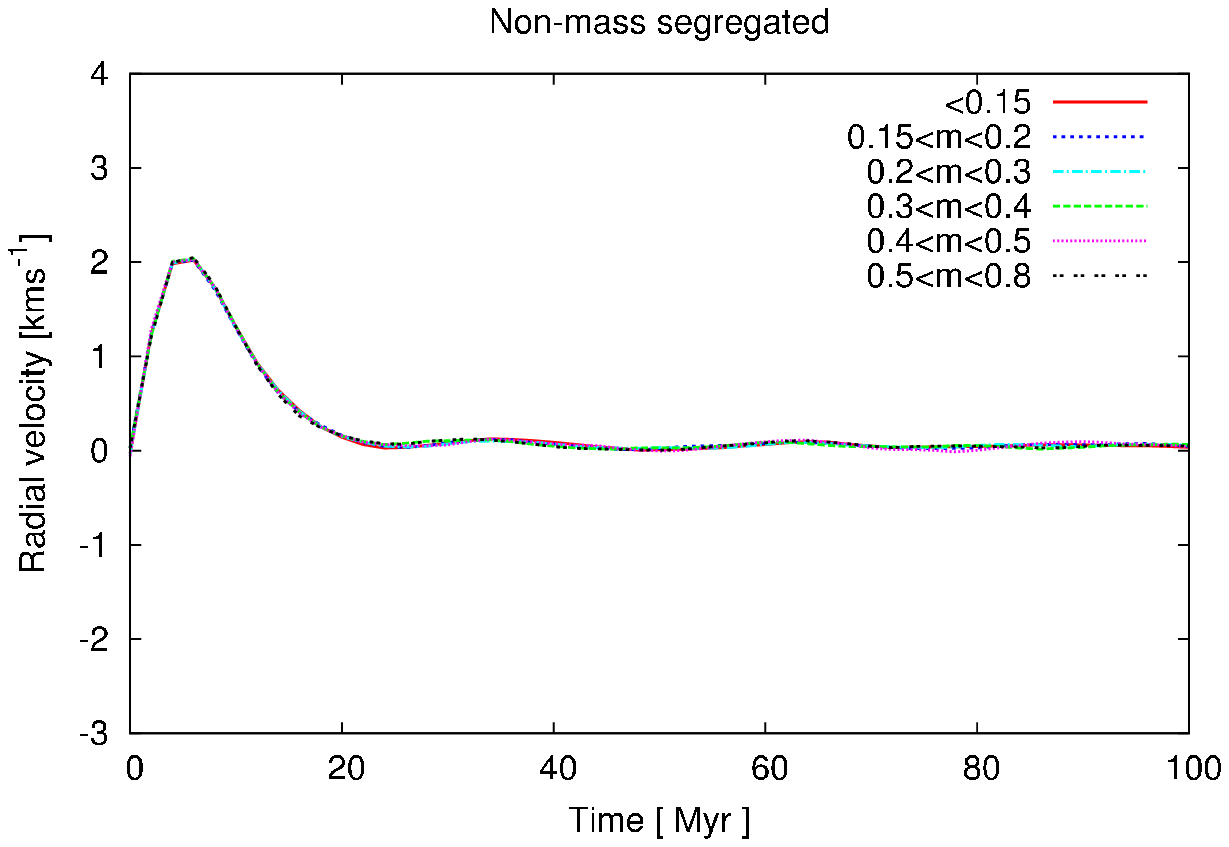}
\includegraphics[width=85mm]{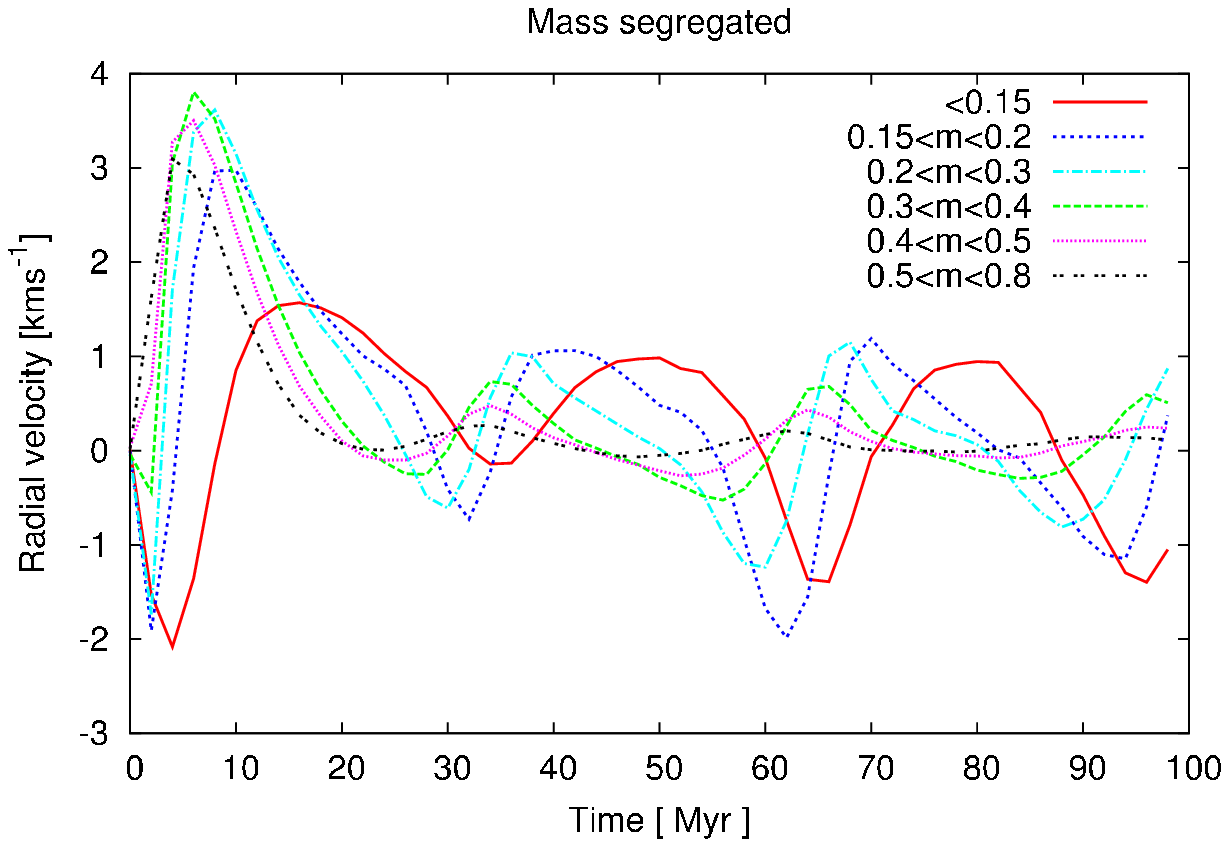}
\caption{The evolution of the mean radial velocity  within the tidal radius of stars in different mass-bins  as in Fig. \ref{v_S9rh10}. The left panel belongs to the tidally-filling non-mass segregated cluster with $r_{h0}=$10 pc and the right one shows the results of a similar cluster but with primordial mass segregation. }
\label{vr_S9rh10}
\end{center}
\end{figure*}

The corresponding evolution of mean radial velocity \footnote{The mean radial velocity is defined as $<\textbf{v}.\textbf{\^{r}}>$} within the tidal radius is demonstrated in Fig. \ref{vr_S9rh10}.  As can be seen all the shells in the not mass-segregated cluster expand equally and after 20 Myr they reach a state of equilibrium. The initial increase in the radial velocity is due to  dilution of the gravitational potential well driven by the early gas expulsion. In fact, in a virialized system, the stars are bound in different orbits and the mean radial velocity of all stars is zero. But, the early ejection of gaseous mass causes a cluster to expand and leads to a positive mean radial velocity  of the  stars. For initially mass-segregated clusters, the mass lost due to the evolution of the massive stars is removed preferentially from the cluster inner region, and the early expansion of the cluster is stronger.

\begin{figure}
\begin{center}
\includegraphics[width=85mm]{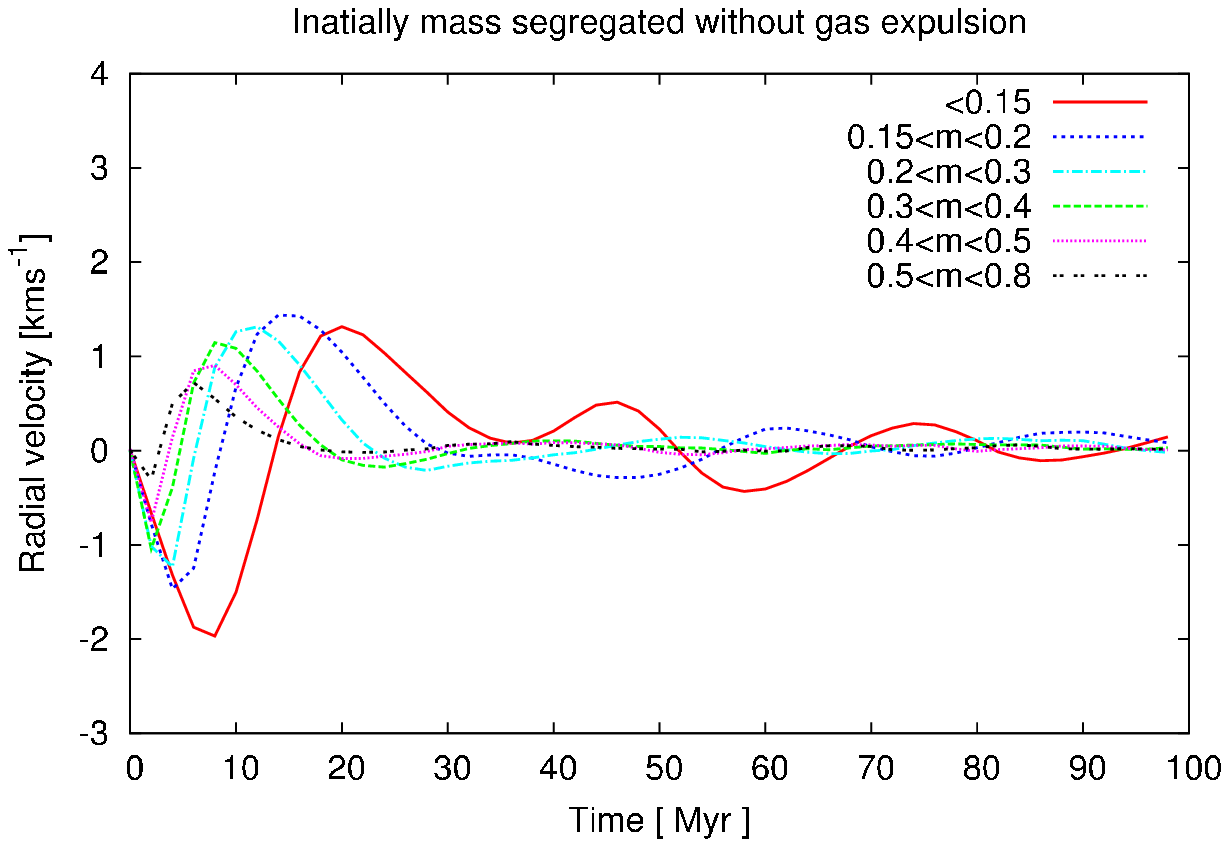}
\includegraphics[width=85mm]{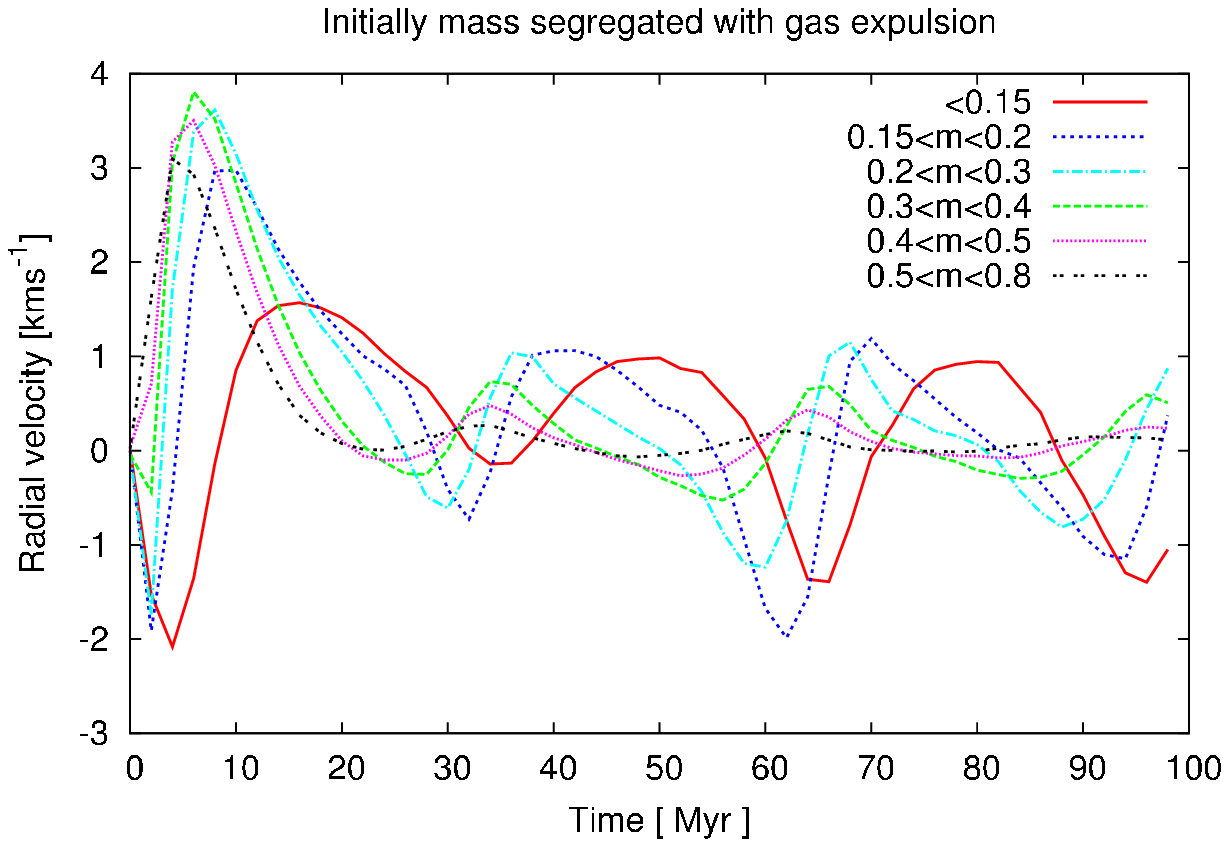}
\caption{The evolution of the mean radial velocity  within the tidal radius of stars in different mass-bins  for two computed initially mass segregated tidally-filling models, 'rh10-S0.9-NoGE' (top: without gas expulsion) and  'rh10-S0-sfe25' (bottom: with gas expulsion). The models are the same,
except that one is w/o gas expulsion. The least massive stars, that initially reside in the outer parts of the cluster,  are least affected by gas expulsion since they have small velocities. }
\label{noge}
\end{center}
\end{figure}

In order to quantify the effect of gas expulsion, we compare the evolution of the mean
radial velocity in  different mass bins for a model without gas expulsion (top panel of Fig. \ref{noge}) with the gas expulsion model
(bottom panel of Fig. \ref{noge}) which is otherwise equal to the case shown in the top panel of Fig. \ref{noge}. Since the least massive stars (that are residing in the outer parts of the cluster) have the smallest velocity dispersion, they are least affected by gas expulsion. The lower mass bins gain a larger mean radial velocity due to gas expulsion leading to an expansion. So, they are more prone to escape from the cluster.

\begin{figure}
\begin{center}
\includegraphics[width=85mm]{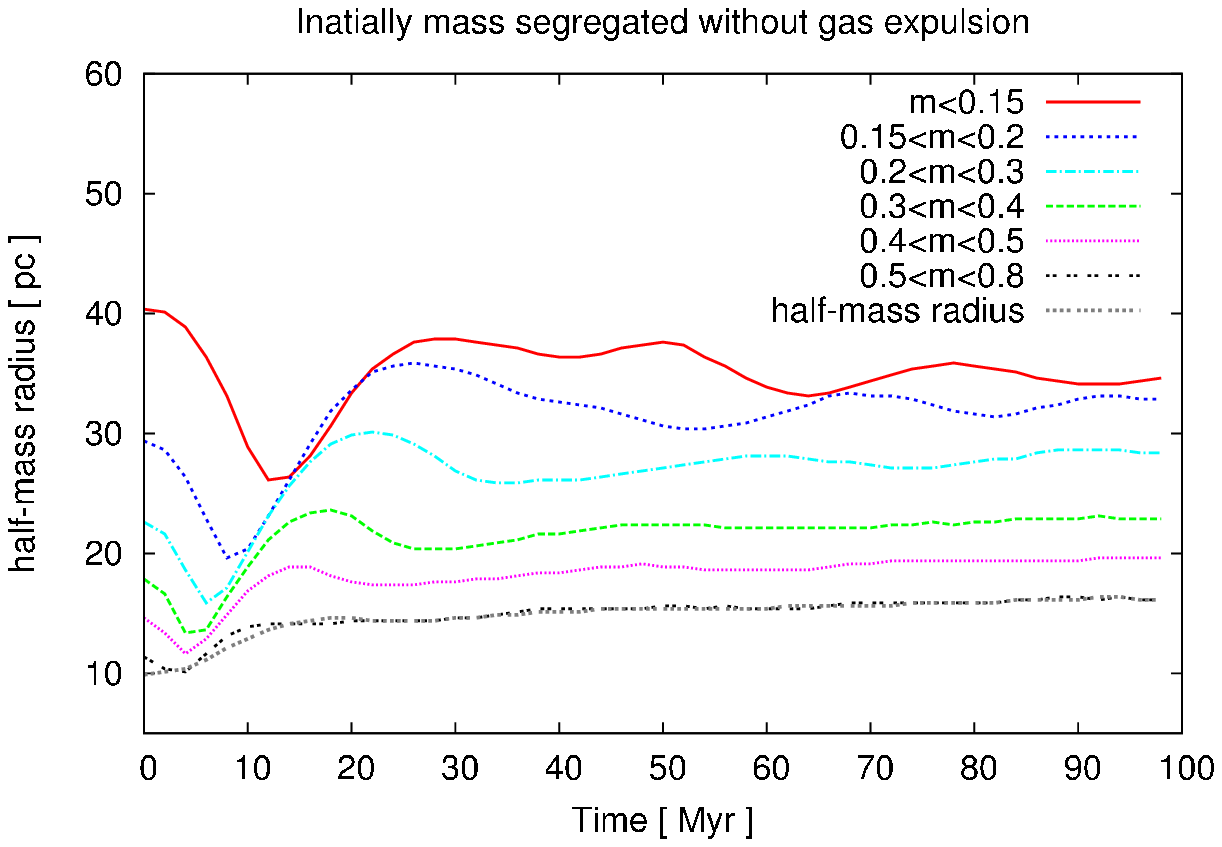}
\includegraphics[width=85mm]{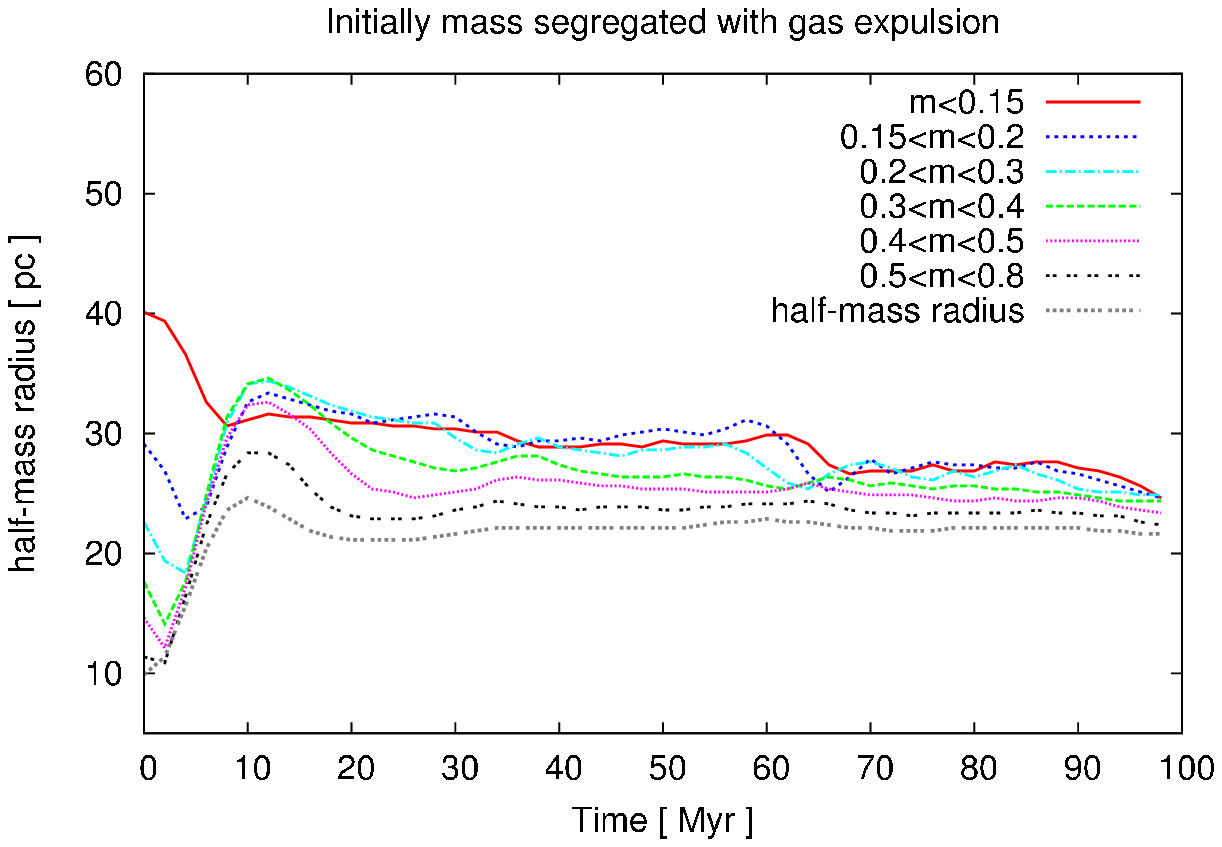}
\caption{The evolution of the half-mass radii of 6 mass-bins chosen within the range 0.08 to 0.80 $M_\odot$, for two computed tidally-filling models as in Fig. \ref{noge}.  The half-mass radius of the cluster is also plotted for comparison.}
\label{r-noge}
\end{center}
\end{figure}

The evolution of the half-mass radii of different mass-bins for both models as a function of time is plotted in Fig. \ref{r-noge}.  Note that in the gas expulsion model, the initial separation of half-mass radii of different mass-bins is reduced as a result of violent relaxation driven by rapid residual gas-expulsion. But, the model without gas expulsion preserves this initial separation and is still mass segregated at $T=100$ Myr.

\section{Conclusion}


We have performed a series of N-body computations studying the impact of expulsion of residual gas
on the stellar masses function slope in star clusters for a wide range of stellar mass, varying the strength of the external tidal field in order to model the large spatial differences of the tidal field expected in the varying proto-MW (Marks \& Kroupa 2010).

Gas expulsion leads to the loss of a substantial fraction of the initially bound mass. A tidal field and mass loss by stellar evolution helps increasing this unbound fraction. Comparing the final masses in the first and last two models in Table 1, one can see that the mass depletion due to gas expulsion can be comparable or even larger than that due to stellar evolution, particularly for clusters subject to a stronger tidal field modelled here as tidally filling clusters.

We have shown that the MFs of non-segregated clusters still resemble their IMFs and that they do not show a strong change of the slope over the whole range of stellar mass over 100 Myr of evolution, because stars of all stellar masses are equally affected by gas expulsion.

In the case of primordially mass segregated clusters, the models divide into two different regimes, depending on the strength of the external tidal field:

\begin{itemize}  
\item Tidally-underfilling models with $r_h/r_t$ values smaller than 0.1 (i.e. when the cluster size is small at first) which keep their MF in the mass range $ m\geq 0.50 M_\odot$ unchanged.
\item Tidally-filling models with $r_h/r_t$ values larger than 0.1, where a large fraction of their low-mass stars are lost and hence they evolve to a comparatively shallow MF.
\end{itemize}

Clusters with $r_h/r_t \leq 0.1$ approximately preserve their MF at the high-mass end (i.e. $ m\geq 0.85 M_\odot$) apart from changes due to stellar evolution, and receive only a very small change in the intermediate stellar mass range ($0.50 M_\odot \leq m \leq 0.85 M_\odot$) even if the SFE is low (SFE $\leq 0.25\%$). But the MF in the low-mass part ($ m\leq 0.5 M_\odot$) changes significantly, because primordial mass segregation leads to a larger impact of stellar evolution on cluster expansion and enhances the preferential loss of less massive stars due to two-body relaxation after post-gas expulsion re-virialisation.

On the other hand, primordially mass-segregated tidally-filling clusters with $r_h/r_t \geq 0.1$ show a strongly depleted mass function in the intermediate stellar mass range ($0.50 M_\odot \leq m \leq 0.85 M_\odot$) as well as in the low-mass part. Also, the slope of the MF at the high-mass end changes slightly.

The mass functions in both cases are quite different from the non-segregated models, especially in the intermediate-mass  and low-mass part: moving towards lower stellar mass, the number of stars per mass interval at first decreases and then starts to increase around $m\simeq0.2 M_\odot$ (i.e. it is not possible to fit the MF with one power-law function).  \emph{If such a behavior in the MF-slope of the low-mass part is observed, this would constitute direct evidence for mass segregation, gas expulsion and a strong tidal field at birth.}

Concerning the dip at the low-mass end of the stellar mass function of fully mass segregated initial clusters, our explanation is that the least massive stars are essentially on radial orbits, since they cannot be moving outwards much given that the cluster is in virial equilibrium. The least massive stars must have the smallest velocity dispersion. So, they probably fall in, or at least, they are least affected by gas expulsion since they have small velocities, while the more massive stars are lost from the cluster due to gas expulsion.

It should be noted that if the outermost stars predominantly move inwards, that is more of an artefact of how the mass-segregated cluster is set up. However, even without the infalling nature, a similar feature would have appeared since the outermost (and lowest mass for a fully mass-segregated cluster) stars are the slowest.

If gas expulsion is a common process in the early stages of star clusters, then by direct measurement of the slope  of the mass function at low stellar mass of very young massive clusters ($\leq 50$ Myr), one can constrain the starting condition of the clusters such as primordial mass segregation. These results confirm the conclusions reached by \cite{Marks08}. Moreover, the shape of the MF can constrain the birth place or the birth size of the cluster, because  the tidally-underfilling clusters have a falling and then a rising trend in the MF-slope in the low-mass end if they are born \emph{initially segregated}.

The models with gas expulsion agree with the general trend that less concentrated clusters show a flattened MF.  We note that the MF-slopes and $c$ values are determined in the models presented here after 100 Myr of evolution, but, since the two-body relaxation time for all models with an altered MF are larger than one-third of the Hubble time (Table 1), the position of the clusters in the $c- \alpha$ plane  will not significantly change over 10 Gyr.

Our models constitute a possible solution to the flattened mass function of the two remote halo GCs, Pal 4 and 14 (Zonoozi et al., in preparation). In our previous publications (Zonoozi et al. 2011, 2014) we have shown that, while two-body relaxation does have an effect on the cluster's stellar mass function, the effect from two-body relaxation is too weak such that the shallow present-day mass function slope measured by Jordi et al. (2009) and Frank et al. (2012) cannot be reproduced with models starting from a canonical Kroupa IMF (2001). We have demonstrated that the observed parameters of globular clusters Pal 4 and Pal 14 (i.e. their observed mass functions, mass segregation, half light radii and masses) can, in principle, be reproduced either by assuming  a very high primordial mass segregation, or by assuming an IMF already depleted in low-mass stars, compared to the canonical Kroupa IMF. We showed that such an initially flattened mass function can be explained by the loss of a good fraction of low-mass stars from an initially mass-segregated embedded cluster across the tidal radius during a violent early residual-gas expulsion phase in agreement with the general findings by Marks \& Kroupa (2010).

Therefore, the flattened IMF of Pal 4 and 14 we have uncovered strengthens the idea that these clusters formed  with a high degree of primordial mass segregation in a stronger tidal field rather than in isolation. I.e. if Pal 4 and 14 have always been on a circular orbit with the present day Galactocentric distance, it is hard to imagine that the tidal field would have played a role during the response of the cluster to gas expulsion. This may have been the case if either the Galactic tidal field has since then evolved \citep{Marks10}, or if the cluster was born within the inner part of our Galaxy,  and then moved to a larger Galactocentric distance on an eccentric orbit, which is more likely than a circular orbit. Or, the cluster birth site was a now detached/disrupted tidal dwarf galaxy (Pawlowski et al 2011). Evidence for this scenario would strengthen if these clusters are found to orbit within the vast polar structure (VPOS, Pawlowski \& Kroupa 2014).

Thus the two most important new results we arrive at are (i) that the observed correlation between concentration and the MF-slope can be accounted for excellently (Fig. 7) and (ii) that the MF of stars ought to have a minimum at about $0.2 M_{\odot}$ (at an age up to $\simeq 100 $ Myr), if GCs formed mass-segregated and suffered gas expulsion in a strongly varying proto-Galactic potential. This confirms the previous conclusions arrived at by \cite{Marks08} and \cite{Marks10}, respectively, yielding an important  if not essential window towards studying the very early phase of MW assembly.  The approach taken here (gas expulsion and tidal field) yields otherwise unobtainable constraints on the very early conditions when the proto-MW was forming. That is, with this approach, we can obtain constraints on a time-resolution of less than a few hundred Myr and on the fluctuations of the very early MW potential.

We emphasize that massive embedded clusters with masses larger than about $10^4\,M_\odot$ revirialise rapidly and within a few Myr, they expand by a factor of 3-4 and loose a small fraction (at most 20~\% in a solar neighborhood tidal field) of their stellar mass as a result of the blow-out within a thermal time-scale (with a gas velocity of 10 km/s) of 67~\% of their original embedding gas  mass (Brinkmann, Banerjee \& Kroupa, in prep.; Banerjee \& Kroupa 2014, 2015). Less massive clusters expand more, take a much longer time to revirialise and are much more damaged losing more than 50~\% of their stars by gas blow-out as shown by Kroupa, Aarseth \& Hurley (2003).  Thus, for those globular clusters, such as studied in this paper, which have a deficit of low-mass stars to be sufficiently damaged by gas blow-out, they must have been also subject to a strong local tidal field for the outer expanded regions of the cluster to be stripped. Only a fraction of all globular clusters suffered this fate (Marks \& Kroupa 2010).

It should be pointed out that our conclusions in the present paper are based on two major assumptions, that is, primordial mass segregation and residual gas expulsion that are questioned by some authors \citep{Kruijssen12, Parker15, Dale12, Dale15, Smith11}.
Concerning the debate whether gas expulsion is  relevant in the formation of star clusters (Sec. 2), we note the high-precession space-astrometry of the GAIA mission is ideally suited to settle this argument, by allowing the kinematical fields around young clusters to be mapped accurately (Kroupa 2005). Simulations with gas expulsion of clusters in the present-day MW need to be done to make predictions of the velocity field for GAIA, because each young cluster should be surrounded by a radially expanding population of stars of similar age as the cluster population. As predicted by Kroupa (2005) the expanding population forms a new moving group stemming from explosive residual gas expulsion that is related to the pre-gas expulsion velocity dispersion of stars in the embedded cluster, and thus to the mass of the re-virilised cluster. This new moving group (MGI according to Kroupa 2005) is present  in addition to the classical moving group (MGII according to Kroupa 2005) stemming from secular evolution. Clustered star formation may thus also affect the kinematics and structural properties of galaxies. For example Kroupa (2002) has shown that the thick disk may be outcome of violent star formation within a thin disk.


\bsp \label{lastpage} \end{document}